\documentclass[aps,pra,twocolumn,superscriptaddress,longbibliography]{revtex4}
\usepackage{graphicx}
\usepackage{dcolumn}
\usepackage{bm}
\usepackage{physics}
\usepackage{amsmath}
\usepackage{amssymb}
\usepackage{array}
\usepackage{soul}
\usepackage{color}
\usepackage{xcolor} 
\newcolumntype{P}[1]{>{\centering\arraybackslash}p{#1}}
\usepackage[colorlinks=true, breaklinks=true, linkcolor=blue, citecolor=blue, urlcolor=blue]{hyperref}

\newcommand{\geneva}{Department of Quantum Matter Physics, University of Geneva, Quai Ernest-Ansermet 24, 1211 Geneva, Switzerland}

\newcommand{\pks}{Max Planck Institute for the Physics of Complex Systems, N\"othnitzer Str.~38, 01187 Dresden, Germany}

\begin{document}

\title{Dissipative generation of currents by nonreciprocal local and global environments}
\date{\today}

\begin{abstract}
We investigate the mechanisms necessary for the stabilization of complex quantum correlations by exploring dissipative couplings to nonreciprocal reservoirs. We analyze the role of locality in the coupling between the environment and the quantum system of interest, as we consider either local couplings throughout the system, or a single global coupling. 
We contrast the results obtained for the two scenarios in which a chain of strongly interacting hardcore bosonic atoms is coupled directly to Markovian kinetic dissipative processes, or experiences effective dissipation through the mediation of the field of a lossy optical cavity.
To investigate the dissipative dynamics of the many-body quantum systems considered we perform numerical simulations employing matrix product states methods. 
We show that by coupling atomic tunneling terms to the global field of a dissipative cavity we can stabilize at long times both finite currents and current-current correlations throughout the atomic chain. This is in contrast to the setup in which dissipation acts directly via local tunneling processes, where currents arise in a narrow region of the system and the current-current correlations are rapidly decaying.
\end{abstract}

\author{Catalin-Mihai Halati}
\affiliation{\pks}
\affiliation{\geneva}
\maketitle

\section{Introduction}

The out-of-equilibrium dynamics of interacting particles stemming from the competition between conservative and dissipative forces has been attracting enormous interest in many fields of physics, ranging from classical \cite{Ramaswamy2010, MarchettiAditi2013, ShankarVitelli2022, BowickRamaswamy2022} to quantum systems \cite{MivehvarRitsch2021, delaTorreSentef2021, SchlawinSentef2022, SiebererDiehl2025}. 
In the field of active matter classical particles are taken out-of-equilibrium by locally breaking detailed balance \cite{Ramaswamy2010, MarchettiAditi2013, VrugtWittkowski2025, VrugtCates2025}, which can lead to novel motional collective behavior \cite{TonerRamaswamy2005}. For example, the active particles can spontaneously choose the direction of motion \cite{EvansMukamel1995, TonerTu1995}, which in equilibrium would be forbidden by the Mermin-Wagner theorem.
Recently, the connection has been made between classical active matter and dissipative quantum systems, where dissipative couplings are employed to induce nonreciprocal processes which can break detailed balance  \cite{SiebererDiehl2025, FruchartVitelli2026, MetelmannClerk2015, KeckFazio2018, YamamotoKawakami2020, AdachiKawaguchi2022, YamagishiObuse2024, KhassehHeyl2025, TakasanKawaguchi2024, LeeClerk2024, ZelleDiehl2024, BrighiNunnenkamp2024, NadolnyBrunelli2025, BelyanskyClerk2025, AntonovLowen2025, HanaiTazai2025, BolechGiamarchi2025, ShiBukov2026, PennervonOppen2025, SoaresSchiro2026}.
A central question in field of quantum active matter relates to how quantum coherence and the long-range quantum correlations can survive under the action of the dissipative processes.

The problem of maintaining quantum coherence and many-body entanglement in the presence of dissipative couplings has received a lot of attention within the field of dissipative engineering \cite{DiehlZoller2008, KrausZoller2008, VerstraeteCirac2009, WeimerBuchler2010, MuellerZoller2012, BernierKollath2013, DuttaCooper2020}. As it turns out, dissipation can be used constructively as an avenue to obtain complex quantum phenomena.
Theoretical efforts have been focused on obtaining steady states with non-trivial nature, e.g.~topological states of fermionic matter \cite{BardynDiehl2012, BardynDiehl2013, BudichDiehl2015}, exotic transport properties \cite{LandiSchaller2022}, or exhibiting dynamical synthetic gauge fields \cite{KollathBrennecke2016, BallantineKeeling2017, HalatiKollath2017, MivehvarPiazza2017}, and on how to engineer the external coupling to control the dynamical properties of a quantum system \cite{PolettiKollath2013, SciollaKollath2015, MacieszczakGarrahan2016, delCampoKim2019, KingMorigi2024, HalatiKollath2025, HalatiJager2025}.
An important concept in dissipative engineering is the decoherence free subspace, i.e.~the subspace spanned by the dark states of the dissipative channels, in which the long-time dynamics of the open quantum system takes place. When the Hamiltonian and the dissipative processes commute the decoherence free subspace contains states which decouple from decay, while in case the Hamiltonian includes many-body couplings competing with the dissipative channels one can extend the notion of the decoherence free subspace to capture also the slowly decaying metastable states \cite{Garcia-RipollCirac2009, PolettiKollath2013, ZanardiCamposVenuti2014, MacieszczakGarrahan2016, JinMa2024, HalatiKollath2025}.
Understanding the quantum states contained in the decoherence free subspace, including the metastable states, is very important when investigating the dynamical behavior of complex correlation. For example, while the superconducting BCS state can be obtain as a dark state of dissipative channels \cite{YiZoller2012}, the two-time current-current correlations are strongly damped in the long-time dissipative dynamics \cite{SciollaKollath2015}.
Thus, strategies are desired in which the coherence of complex quantum correlations in dissipative scenarios is maintained at long times.

In this work, we investigate the long-time dynamics of quantum correlations in the presence of nonreciprocal dissipative couplings for a one-dimensional quantum system of strongly interacting bosonic atoms. We contrast numerical exact results obtained with methods based on matrix product states for systems in which the dissipative processes are either local, or mediated globally by a bosonic field.
We show that finite atomic currents can be stabilized by employing local nonreciprocal dissipative kinetic processes. However, the current-current correlations are either absent or quickly decay with distance, signaling a lack of long-range quantum coherence in the long-time state of the system.
In order to induce current-current correlations we couple the interacting atoms to a dissipative optical cavity. By employing a transverse pump laser beam we couple the photon field to atomic tunneling terms. We show that in the atoms-cavity setup, in which the atoms effectively experience dissipation via a nonreciprocal global reservoir, both atomic currents and the corresponding quantum correlations can be stabilized at long times.

\begin{figure}[!hbtp]
\centering
\includegraphics[width=0.45\textwidth]{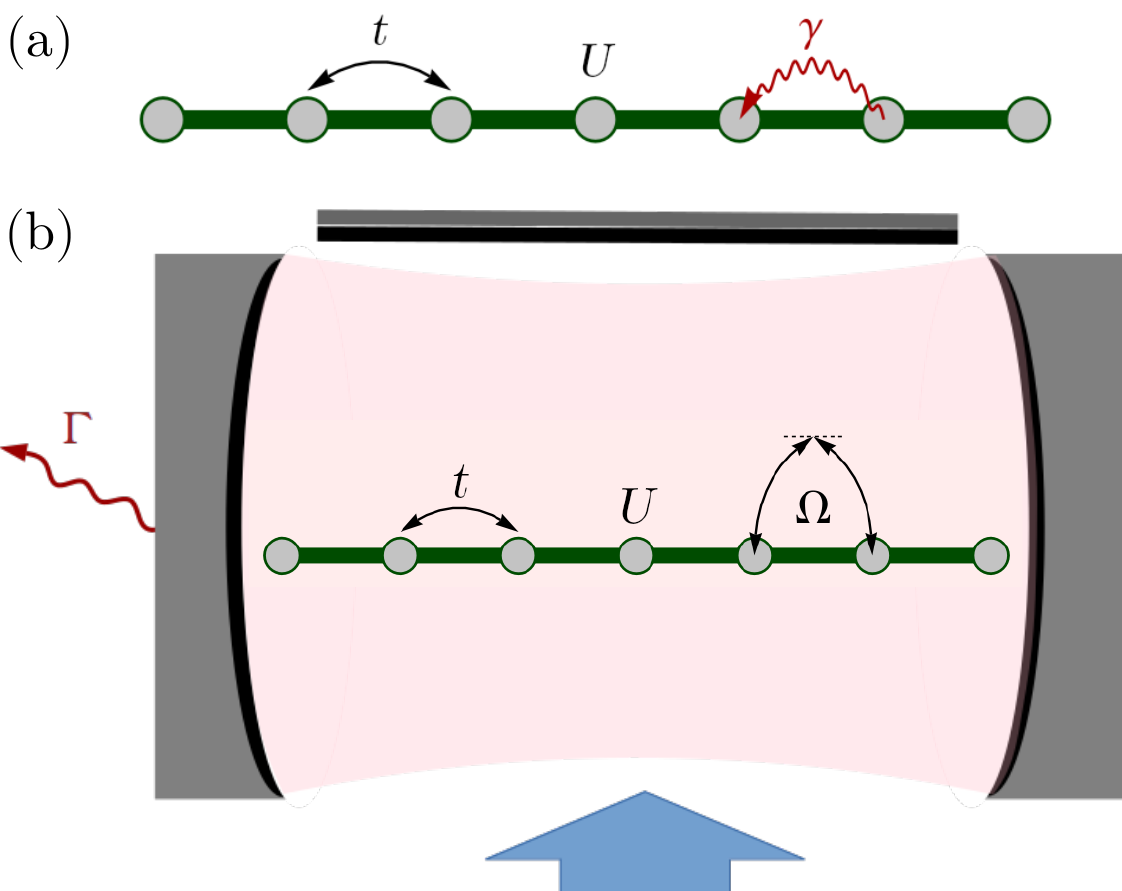}
\caption{Sketches of the models: (a) A one-dimensional chain of interacting bosonic atoms under the action of directional kinetic dissipation. The coherent tunneling processes have the amplitude $t$, the repulsive on-site interactions strength $U$ and the dissipative rate is $\gamma$.
(b) A one-dimensional chain of interacting bosonic atoms coupled to the field of an optical cavity. The atoms-cavity coupling is realized with the help of a retroreflected transverse pump beam and the strength of the coupling is $\Omega$.
Photons are leaking out of the cavity with the dissipation strength $\Gamma$.
}
\label{fig:sketch}
\end{figure}

There are several motivations for investigating the dynamics of currents in many-body quantum systems and devising frameworks for their generation and measurement. 
For example, in transport measurements like the Hall effect one monitors the transverse current resulting upon the application of a potential. This measurement led to the identification of the exotic quantum Hall states \cite{vonKlitzing1986, StormerGossard1999, XiaoNiu2010}. 
The Hall response has recently become accessible also for ultracold atoms, where it has been measured in weakly interacting gases \cite{AidelsburgerGoldman2015, ChalopinNascimbene2020} and ladder systems \cite{ZhouFallani2023, ZhouFallani2025}. While a complete understanding of the Hall response for interacting systems is an open questions, studies for ladders have shown that it is a sensitive probe of the underlying many-body phase diagram of chiral phases \cite{GreschnerGiamarchi2019, BuserGiamarchi2021, CitroOrignac2025, HalatiGiamarchi2025, HalatiGiamarchi2025b}.
Furthermore, the interplay of orbital effects stemming from artificial magnetic fields and the coupling to cavity fields can lead to non-equilibrium steady states characterized by persistent currents \cite{KollathBrennecke2016, SheikhanKollath2016, ZhengCooper2016, BallantineKeeling2017, HalatiKollath2017, HalatiKollath2019, MivehvarPiazza2017, ColellaRitsch2019}.

The plan of the paper is as follows, in Sec.~\ref{sec:setup} we describe the systems we investigate, with the atomic model under dissipative kinetic processes shown in Sec.~\ref{sec:atomic_kinetic_model} and the atoms-cavity model in Sec.~\ref{sec:cavity_model}. The parameter regimes considered in this work are detailed in Sec.~\ref{sec:params}.
In Sec.~\ref{sec:currents} we define the current operators and the associated correlations on which we focus our discussions.
In Sec.~\ref{sec:numerics} we briefly present the numerical methods based on matrix product states employed.
We present our results in Sec.~\ref{sec:results}, with Sec.~\ref{sec:results_atoms} focusing on the results for the atomic model under kinetic dissipation and Sec.~\ref{sec:results_cavity} focusing on the results for the atoms-cavity model and the comparison between the two scenarios.
We conclude with Sec.~\ref{sec:conclusions}.

\section{Setups and Models\label{sec:setup}}

We consider interacting bosonic atoms confined to a one-dimensional optical lattice, under the action of nonreciprocal dissipative couplings aimed at inducing atomic currents.
In the first model, described in Sec.~\ref{sec:atomic_kinetic_model}, dissipation acts directly on the atoms in the form of local tunneling processes. In the second model, described in Sec.~\ref{sec:cavity_model}, the field of an optical cavity is coupled to atomic tunneling terms, with photon losses occurring via the mirrors of the cavity. In this scenario, the atoms experience an effective global dissipation.

\subsection{Kinetic dissipation \label{sec:atomic_kinetic_model}}

The first setup that we investigate is a one-dimensional Bose-Hubbard model with directional tunneling dissipative processes, as sketched in Fig.~\ref{fig:sketch}(a). We consider that the dissipative processes are Markovian, thus, the dynamics of the atomic density matrix $\rho$ is given by the following Gorini-Kossakowski-Sudarshan-Lindblad (GKSL) master equation  \cite{GoriniSudarshan1976, Lindblad1976, CarmichaelBook, BreuerPetruccione2002}
\begin{align}
\label{eq:Lindblad_atoms}
\pdv{t} \hat{\rho} &= -\frac{i}{\hbar} \left[ \hat{H}_\text{BH}, \hat{\rho} \right]  \\
&+ 	\frac{\gamma}{2}\sum_{j=1}^{L-1}\left(2\hat{L}_j\hat{\rho} \hat{L}_j^\dagger-\hat{L}_j^\dagger \hat{L}_j \hat{\rho}-\hat{\rho}\hat{L}_j^\dagger\hat{L}_j  \right), \nonumber 
\end{align}
with the dissipative processes of strength $\gamma$ given by the jump operator $\hat{L}_j=\hat{b}_{j}^\dagger \hat{b}_{j+1}$. 
The jump operator has a similar form to the ones used in other dissipative engineering proposals, e.g.~in Refs.~\cite{DiehlZoller2008, KrausZoller2008, KeckFazio2018, YamamotoKawakami2020, BolechGiamarchi2025}.
Such a dissipative process can be, in principle, realized experimentally by employing two-photon Raman transitions making use of lossy photon modes acting on each bond of the chain. 
By realizing couplings of the form $\omega(a_{j}^\dagger b^\dagger_{j} b_{j+1} + \text{H.c.})$, with $a_{j}^\dagger$ the creation operator for a photon mode labeled by the lattice site and $\omega$ is the effective strength of the Raman transition, and considering very short lifetimes of the photon modes, one can effectively obtain the dissipative processes from Eq.~(\ref{eq:Lindblad_atoms}) \cite{JagerBetzholz2022, BolechGiamarchi2025, SchmitJager2026}.
The coherent part of the dynamics is described by the Bose-Hubbard Hamiltonian 
\begin{align} 
\label{eq:Hamiltonian_BH}
\hat{H}_{\text{BH}}&= \hat{H}_{\text{int}}+\hat{H}_{\text{kin}}\\
\hat{H}_{\text{int}}&=\frac{U}{2} \sum_{j=1}^{L} \hat{n}_{j}(\hat{n}_{j}-1),\nonumber \\
\hat{H}_{\text{kin}}&=-t \sum_{j=1}^{L-1} \left(\hat{b}_{j}^\dagger \hat{b}_{j+1} + \text{H.c.}\right), \nonumber
\end{align}
where $t$ is the amplitude of the atomic tunneling and the repulsive on-site interactions have the strength $U$. $\hat{b}_{j}$ and $\hat{b}_{j}^\dagger$ are bosonic operators and the local density operator is $\hat{n}_{j}=\hat{b}_{j}^\dagger \hat{b}_{j}$. 
We consider $N$ particles on a one-dimensional chain of length $L$. 
In the following, we refer to Eqs.~(\ref{eq:Lindblad_atoms})-(\ref{eq:Hamiltonian_BH}) as the \emph{local atomic model}.

\subsection{Cavity mediated dissipation \label{sec:cavity_model}}

The second setup is a hybrid system consisting in a one-dimensional lattice of interacting bosonic atoms inside a high finesse optical cavity, transversely pumped with a standing-wave laser beam and experiencing photon losses through the cavity mirrors, sketched in Fig.~\ref{fig:sketch}(b). 
Ultracold atoms in optical cavities have emerged as one of the main platforms to engineer dissipative couplings and long-range interactions \cite{RitschEsslinger2013, MivehvarRitsch2021}.
The density matrix of the coupled atoms-cavity degrees of freedom, $\mu$, evolves under the following GKSL master equation \cite{GoriniSudarshan1976, Lindblad1976, CarmichaelBook, BreuerPetruccione2002, RitschEsslinger2013, MivehvarRitsch2021}
\begin{align}
\label{eq:Lindblad}
& \pdv{t} \hat{\mu} = -\frac{i}{\hbar} \left[ \hat{H}_\text{ac}, \hat{\mu} \right] + \frac{\Gamma}{2}\left(2\hat{a}\hat{\mu} \hat{a}^\dagger-\hat{a}^\dagger \hat{a} \hat{\rho}-\hat{\mu} \hat{a}^\dagger \hat{a}\right), 
\end{align}
where $\Gamma$ describes the strength of the photon losses through the mirrors and the atoms-cavity Hamiltonian is given by
\begin{align} 
\label{eq:Hamiltonian_ac}
& \hat{H}_\text{ac}=\hat{H}_{\text{BH}}+\hbar\delta\hat{a}^\dagger \hat{a} -\hbar\Omega \sum_{j=1}^{L-1} \left(\hat{a}^\dagger \hat{b}_{j}^\dagger \hat{b}_{j+1} + \text{H.c.}\right),
\end{align}
where besides the Bose-Hubbard Hamiltonian $\hat{H}_{\text{BH}}$ we also have the energy of the cavity mode and the atoms-cavity coupling terms.
The coupling is realized by two-photon Raman transitions employing a cavity photon and a pump photon via an intermediate excited state and the application of a linear potential \cite{ZhengCooper2016, LaflammeZoller2017}. This coupling has been shown to give rise to atomic currents in non-interacting and small systems \cite{ZhengCooper2016, LaflammeZoller2017}.
$\delta$ is the detuning of the cavity with respect to the pump frequency and the atoms-cavity coupling has an effective strength $\Omega$.
We note that the coupling of the cavity mode to tunneling terms can arise also in other geometries and gives rise to non-trivial steady states of the coupled system \cite{MaschlerRitsch2008, HabibianMorigi2013, Caballero-BenitezMekhov2015, Caballero-BenitezMekhov2016, KollathBrennecke2016, SheikhanKollath2016, HalatiKollath2017, ChandaZakrzewski2021, ChandaMorigi2022}.
The cavity photon losses represent a nonreciprocal dissipative process, as the cavity does not absorb photons form the electromagnetic environment to which the photons are lost. Furthermore, the cavity is coupled to a non-Hermitian atomic operator, thus, also the atoms are effectively coupled to a nonreciprocal environment.
In the following sections, we refer to Eqs.~(\ref{eq:Lindblad})-(\ref{eq:Hamiltonian_ac}) as the \emph{global atom-cavity model}.

\subsection{Parameter regimes \label{sec:params}}

In this work we compare results for the quantum dynamics of the two models described in Sec.~\ref{sec:atomic_kinetic_model}, the local atomic model Eqs.~(\ref{eq:Lindblad_atoms})-(\ref{eq:Hamiltonian_BH}), and Sec.~\ref{sec:cavity_model}, the global atom-cavity model Eqs.~(\ref{eq:Lindblad})-(\ref{eq:Hamiltonian_ac}), with a focus on the dissipative effects on the atomic currents and the associated current-current correlations. Thus, it is helpful to contrast the nature of the two setups, underlying the main differences.
The local atomic model given in Eqs.~(\ref{eq:Lindblad_atoms})-(\ref{eq:Hamiltonian_BH}) corresponds to the atoms experiencing \emph{local} nonreciprocal Markovian environments, while in the atom-cavity model, given in Eqs.~(\ref{eq:Lindblad})-(\ref{eq:Hamiltonian_ac}), the atoms are coupled to a \emph{global} quantum field which is subjected to nonreciprocal losses. 
We emphasize that atomic model, Eqs.~(\ref{eq:Lindblad_atoms})-(\ref{eq:Hamiltonian_BH}), does not represent an effective atoms-only description of the atom-cavity model.
Finding the proper atom-only description of the atom-cavity model is beyond the scope of our work, however, as we detail in the following, it is useful for understanding the effective processes which the atoms experience.

By adiabatically eliminating the cavity degree of freedom \cite{RitschEsslinger2013, MivehvarRitsch2021} one can obtain a GKSL description with a Hamiltonian term $\propto \frac{\hbar\delta\Omega^2}{\delta^2+\Gamma^2/4} \mathcal{O}^\dagger \mathcal{O}$, describing the long-range interactions induced by the cavity, and an effective Markovian dissipation with the jump operator $\mathcal{O}$ with strength $\frac{\hbar\Gamma\Omega^2}{2\delta^2+\Gamma^2/2}$, where $\mathcal{O}$ is the operator to which the cavity is coupled to. 
In our case $\mathcal{O}=\sum_{j=1}^{L-1}  \hat{b}_{j}^\dagger \hat{b}_{j+1}$, which is not an Hermitian operator. Thus, the atoms effectively experience  nonreciprocal dissipation.
However, such a description is valid only in the limit in which the cavity energy scales $\hbar\delta$ and $\hbar\Gamma$ are much larger than the atoms-cavity coupling and the atomic energies, beyond this adiabatic limit deviations appear \cite{DamanetKeeling2019, HalatiKollath2020, BezvershenkoRosch2021, HalatiKollath2022, LinkDaley2022, JagerBetzholz2022, HalatiKollath2025, TolleHalati2025, HalatiJager2025, SchmitJager2026, OrsoDeuar2025}.
In particular, by adding corrections to $\mathcal{O}$ one can extend the range of validity of the atom-only effective GKSL description \cite{JagerBetzholz2022, SchmitJager2026}, or consider descriptions in which the atoms are coupled to a non-Markovian environment \cite{LinkDaley2022, DebeckerDamanet2024, DebeckerDamanet2025}.
We note that as we employ numerical exact simulations we can capture the dynamics of the quantum correlations regardless of the parameter regime  \cite{HalatiKollath2020, BezvershenkoRosch2021, HalatiKollath2022, HalatiKollath2025}, as detailed in Sec.~\ref{sec:numerics}.
To summarize, in the strongly dissipative regime, $\Gamma>\delta$, we can view the coupling to the cavity, Eqs.~(\ref{eq:Lindblad})-(\ref{eq:Hamiltonian_ac}), as the coupling to a \emph{global} environment, with the main contribution given by a dissipative channel corresponding to $\mathcal{O}=\sum_j  \hat{b}_{j}^\dagger \hat{b}_{j+1}$, but with non-negligible memory and retardation effects.
The fact that $\mathcal{O}=\sum_j  \hat{b}_{j}^\dagger \hat{b}_{j+1}$ is a global operator leads to an important distinction compared to the model with locally acting dissipation, Eqs.~(\ref{eq:Lindblad_atoms})-(\ref{eq:Hamiltonian_BH}), as the decoherence free subspace will have a much larger dimension, leading to the possibility of stabilizing complex long-range correlations at long times, as we explore in our results, Sec.~\ref{sec:results}.

We consider strong on-site interactions for the bosonic atoms corresponding to the hard-core limit, $U\to\infty$. As we perform simulations for a finite size system, this will prevent the atoms to accumulate all on the first site of the chain under the action of the emerging currents. 
In this case, under just the action of the dissipative terms in Eq.~(\ref{eq:Lindblad_atoms}) the steady state will be the state $\ket{1\dots 10\dots0}$, with one particle per site for the first $N$ sites. However, once the coherent Hamiltonian is taken into account deviations from this steady state can occur.

The main parameters used in this work are as follows. We consider a finite size chain with $L=32$ sites of hardcore bosons at quarter filling, i.e.~$N=8$ particles.
As we aim to understand the dissipative effects, for the local atomic model with kinetic dissipation, Eqs.~(\ref{eq:Lindblad_atoms})-(\ref{eq:Hamiltonian_BH}), we take the range of the dissipation strength from being comparable to the kinetic energy to being the dominant process, with $0.5\leq \hbar\gamma/t \leq 8$.
Similarly, for the global atom-cavity model, Eqs.~(\ref{eq:Lindblad})-(\ref{eq:Hamiltonian_ac}), we want the dominant effects stemming from the coupling to the cavity to be due to the global dissipative coupling to an environment and not due to the long-range coherent interactions, thus, we take $\delta<\Gamma$. 
In particular, we use $\hbar\Omega/t=2.5$, $\hbar\delta/t=5$ and $10\leq \hbar\Gamma/t \leq 60$. 
For reference, this would result in an effective coupling strength of the long-range interactions varying in between $0.625 t\geq\frac{\hbar\delta\Omega^2}{\delta^2+\Gamma^2/4}\geq 0.033 t$, and the effective dissipation strength varying in between $0.625 t\geq\frac{\hbar\Gamma\Omega^2}{2\delta^2+\Gamma^2/2}\geq0.202 t$.
One should note that both the long-range interaction term and the global dissipative terms scale $\propto L^2$, while the kinetic Hamiltonian term scales $\propto L$, thus, even though the values of the effective couplings are smaller than $t$, the cavity induced processes are still the dominant ones.

For both systems we consider the same protocol, the initial state consists in the ground state of the hardcore bosonic atoms, at time $\tau=0$ we quench the kinetic dissipation, or the coupling to the cavity, respectively. We calculate the time-evolution of the systems following the quench up to times approaching the steady state.

\section{Current Operators \label{sec:currents}}

We define the current operators based on the continuity equation, which relates the time-evolution of the local density to the currents flowing in and out
\begin{align} 
\label{eq:continuity}
& \frac{\partial}{\partial t} \hat{n}_l= \sum \hat{J}_{in} - \sum\hat{J}_{out}.
\end{align}
In particular, we have contributions from the Bose-Hubbard terms, which are our main focus in this work, but also contributions from the dissipative part \cite{HovhannisyanImparato2019} for the local atomic model given in Eqs.~(\ref{eq:Lindblad_atoms})-(\ref{eq:Hamiltonian_BH}) and from the cavity coupling for the global atom-cavity model given in Eqs.~(\ref{eq:Lindblad})-(\ref{eq:Hamiltonian_ac}).
While in the steady state the local densities do not evolve, $\frac{\partial}{\partial t} \hat{n}_l=0$, our results show that the individual contributions are not necessarily independently vanishing.

\subsection{Bose-Hubbard currents}

For the Bose-Hubbard Hamiltonian, Eq.~(\ref{eq:Hamiltonian_BH}), we obtain the following continuity equation
\begin{align} 
\label{eq:continuity_BH}
 \frac{\partial}{\partial t} \hat{n}_l&=i[\hat{H}_\text{BH},\hat{n}_l] \\
 =&-it \left(\hat{b}_{l-1}^\dagger \hat{b}_{l}-\hat{b}_{l}^\dagger \hat{b}_{l-1} \right)+it \left(\hat{b}_{l}^\dagger \hat{b}_{l+1}-\hat{b}_{l+1}^\dagger \hat{b}_{l} \right). \nonumber
\end{align}
This allows to define the following expression for the current operator
\begin{align} 
\label{eq:current_BH}
 \hat{J}_l=&-it \left(\hat{b}_{l}^\dagger \hat{b}_{l+1}-\hat{b}_{l+1}^\dagger \hat{b}_{l} \right), 
\end{align}
with its expectation value being related to the imaginary part of single particle correlations
\begin{align} 
\label{eq:current_BH_av}
 \langle\hat{J}_l\rangle=2t\Im\left\langle\hat{b}_{l}^\dagger \hat{b}_{l+1}\right\rangle. 
\end{align}

The associated long-range current-current correlations can be written as
\begin{align} 
\label{eq:current_current}
 \hat{C}_{JJ}(l,d)=\frac{1}{t^2}\hat{J}_l\hat{J}_{l+1+d}, 
\end{align} 
with the expectation value given in terms of four-point correlations functions
\begin{align} 
\label{eq:current_current_av}
 \langle \hat{C}_{JJ} \rangle(l,d)=&2\Re\left\langle\hat{b}_{l} \hat{b}^\dagger_{l+1}\hat{b}_{l+d+1}^\dagger \hat{b}_{l+d+2}\right\rangle \\
 -&2\Re\left\langle\hat{b}^\dagger_{l} \hat{b}_{l+1}\hat{b}_{l+d+1}^\dagger \hat{b}_{l+d+2}\right\rangle.\nonumber
\end{align}

\subsection{Contributions to the currents from kinetic dissipation}

For the case in which the dissipative couplings to the environment act directly on the atomic tunneling, as in the local atomic model given in Eqs.~(\ref{eq:Lindblad_atoms})-(\ref{eq:Hamiltonian_BH}), the time-evolution of the local densities is changed as in the following
\begin{align} 
\label{eq:continuity_kin_dis}
 \frac{\partial}{\partial t} \hat{n}_l&=i[\hat{H}_\text{BH},\hat{n}_l]+\frac{\gamma}{2}\sum_j \left(2\hat{L}_j^\dagger \hat{n}_l \hat{L}_j-\hat{L}_j^\dagger \hat{L}_j \hat{n}_l-\hat{n}_l \hat{L}_j^\dagger \hat{L}_j  \right)\nonumber \\
 =&\hat{J}_{l-1} -\gamma \hat{b}^\dagger_{l} \hat{b}_{l}\hat{b}_{l-1} \hat{b}_{l-1}^\dagger - \hat{J}_l + \gamma \hat{b}^\dagger_{l+1} \hat{b}_{l+1}\hat{b}_{l} \hat{b}_{l}^\dagger.
\end{align}
Thus, we can identify the following dissipative contribution to the currents 
\begin{align} 
\label{eq:current_kin_dis}
 \hat{K}^\gamma_l=&-\gamma \hat{b}^\dagger_{l+1} \hat{b}_{l+1}\hat{b}_{l} \hat{b}_{l}^\dagger, 
\end{align}
which for hardcore bosons has the following dependence on the density-density correlations between neighboring sites and the local densities
\begin{align} 
\label{eq:current_kin_dis_av}
\langle\hat{K}^\gamma_l\rangle=&\gamma \left(\langle\hat{n}_l\hat{n}_{l+1}\rangle-\langle\hat{n}_{l+1}\rangle\right).
\end{align}

\subsection{Cavity contributions to the currents}

For the global atom-cavity model in which the cavity field couples to tunneling terms, Eqs.~(\ref{eq:Lindblad})-(\ref{eq:Hamiltonian_ac}), while we do not have contributions from the dissipative part as the losses act only on the cavity field, we do have contributions from the atoms-cavity coupling term
\begin{align} 
\label{eq:continuity_kin_cav}
 \frac{\partial}{\partial t} \hat{n}_l=&i[\hat{H}_\text{ac},\hat{n}_l]\\
 =&\hat{J}_{l-1} -i\hbar\Omega \left(\hat{a}^\dagger\hat{b}_{l-1}^\dagger \hat{b}_{l}-\hat{a}\hat{b}_{l}^\dagger \hat{b}_{l-1} \right) \nonumber\\
 &- \hat{J}_l + i\hbar\Omega \left(\hat{a}^\dagger\hat{b}_{l}^\dagger \hat{b}_{l+1}-\hat{a}\hat{b}_{l+1}^\dagger \hat{b}_{l} \right). \nonumber
\end{align}
We observe that the terms coming from the atoms-cavity coupling stem from the correlated motion of the atoms with the creation, or annihilation, of a cavity photon 
\begin{align} 
\label{eq:current_kin_cav}
 \hat{K}^\text{ac}_l=&-i\hbar\Omega \left(\hat{a}^\dagger\hat{b}_{l}^\dagger \hat{b}_{l+1}-\hat{a}\hat{b}_{l+1}^\dagger \hat{b}_{l} \right), 
\end{align}
with the corresponding expectation value
\begin{align} 
\label{eq:current_kin_cav_av}
 \langle\hat{K}^\text{ac}_l\rangle=&2\hbar\Omega \Im\left\langle\hat{a}^\dagger\hat{b}_{l}^\dagger \hat{b}_{l+1}\right\rangle. 
\end{align}

\section{Numerical Methods \label{sec:numerics}}

Our results for the numerically exact quantum dynamics of the models presented in Sec.~\ref{sec:setup} are based on matrix product states (MPS) techniques \cite{Schollwoeck2011, PaeckelHubig2019}.
We deal with the dissipative couplings by employing the stochastic unravelling of the master equation with quantum trajectories \cite{DalibardMolmer1992, GardinerZoller1992, Daley2014}.
In this approach one simulates wavefunctions, i.e.~quantum trajectories, instead of density matrices, however, with the additional complexity of having to perform the Monte-Carlo average over all sampled trajectories.
For the atomic model with short-range interactions, Eqs.~(\ref{eq:Lindblad_atoms})-(\ref{eq:Hamiltonian_BH}), we perform the time-evolution making use of the quasi-exact time-dependent variational matrix product state (tMPS) based on the Trotter-Suzuki decomposition of the time evolution propagator \cite{WhiteFeiguin2004, DaleyVidal2004, Schollwoeck2011}.
The numerical results for the global atom-cavity model, Eqs.~(\ref{eq:Lindblad})-(\ref{eq:Hamiltonian_ac}), which contains the global range couplings, were obtained with an implementation of the two-site version of the time-dependent variational principle approach (TDVP) based on matrix product states  \cite{HaegemanVerstraete2011, HaegemanVerstraete2016}.
Our methods have been implemented with the use of the ITensor Library \cite{FishmanStoudenmire2020}. 
For both models the hardcore limit of the bosonic atoms is implemented by considering a local Hilbert space of two states.

For the atomic model with kinetic dissipation, Eqs.~(\ref{eq:Lindblad_atoms})-(\ref{eq:Hamiltonian_BH}), the convergence of the presented results was ensured by the following convergence parameters: a maximal bond dimension of $250$ states, which ensured a truncation error of at most $2\times10^{-10}$ at the final time, and a time-step of $d\tau t/\hbar=4\times10^{-3}$. We average the results over $5000$ quantum trajectories.
We present the results with the error bars based on the standard deviation of the Monte-Carlo sampling, when the error bars are absent the stochastic errors are smaller than the size of the symbols employed.

For the global atom-cavity model, Eqs.~(\ref{eq:Lindblad})-(\ref{eq:Hamiltonian_ac}), the convergence of the presented results was ensured by the following convergence parameters: a maximal bond dimension of $400$ states, which ensured a truncation error of at most $10^{-9}$ at the final time, a time-step of $d\tau t/\hbar=4\times10^{-3}$, and the adaptive cutoff of the local Hilbert space of the photonic mode ranged between $N_\text{pho}=35$ and $N_\text{pho}=7$ \cite{HalatiKollath2020b, TolleHalati2025c}. We average the results over $500$ quantum trajectories.
As for the other model, we present the results with the error bars based on the standard deviation of the Monte-Carlo sampling, when the error bars are absent the stochastic errors are smaller than the size of the symbols employed.
Additional details regarding the TDVP implementation based on MPS for the atoms-cavity dissipative model, together with convergence benchmarks and comparisons with implementations based on the Trotter-Suzuki decomposition and swap gates \cite{WallRey2016, HalatiKollath2020b}, can be found in Ref.~\cite{TolleHalati2025c}.

\section{Results \label{sec:results}}

In the following, we present the results for the dissipative dynamics for the two models we consider. We focus our discussion the behavior of current observables and the associated current-current correlations (see Sec.~\ref{sec:currents}), however, we show results also for other observables in order to get a deeper understanding of the resulting dynamical behavior. We first discuss in Sec.~\ref{sec:results_atoms} the results for the system of hardcore bosonic atoms experiencing kinetic dissipation [sketched in Fig.~\ref{fig:sketch}(a)], Eqs.~(\ref{eq:Lindblad_atoms})-(\ref{eq:Hamiltonian_BH}). We contrast these results with the ones obtained for the hardcore bosonic atoms coupled to a dissipative cavity [sketched in Fig.~\ref{fig:sketch}(b)], Eqs.~(\ref{eq:Lindblad})-(\ref{eq:Hamiltonian_ac}), in Sec.~\ref{sec:results_cavity}.

\subsection{Dynamics of hardcore bosons with kinetic dissipation \label{sec:results_atoms}}

\begin{figure}[!hbtp]
\centering
\includegraphics[width=0.4\textwidth]{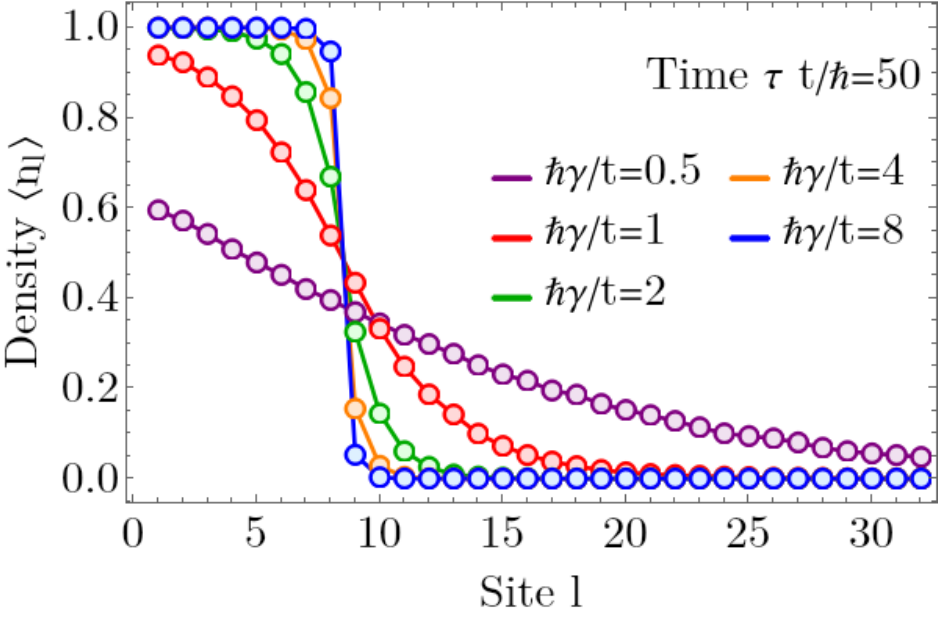}
\caption{The local density profile, $\langle \hat{n}_l \rangle$, at time $\tau t/\hbar=50$, for the local atomic model with kinetic dissipation, Eqs.~(\ref{eq:Lindblad_atoms})-(\ref{eq:Hamiltonian_BH}), and different values of the dissipation strength, $\hbar\gamma/t\in\{0.5,1,2,4,8\}$. We consider $L=32$ sites and $N=8$ particles.
}
\label{fig:density}
\end{figure}

\begin{figure}[!hbtp]
\centering
\includegraphics[width=0.48\textwidth]{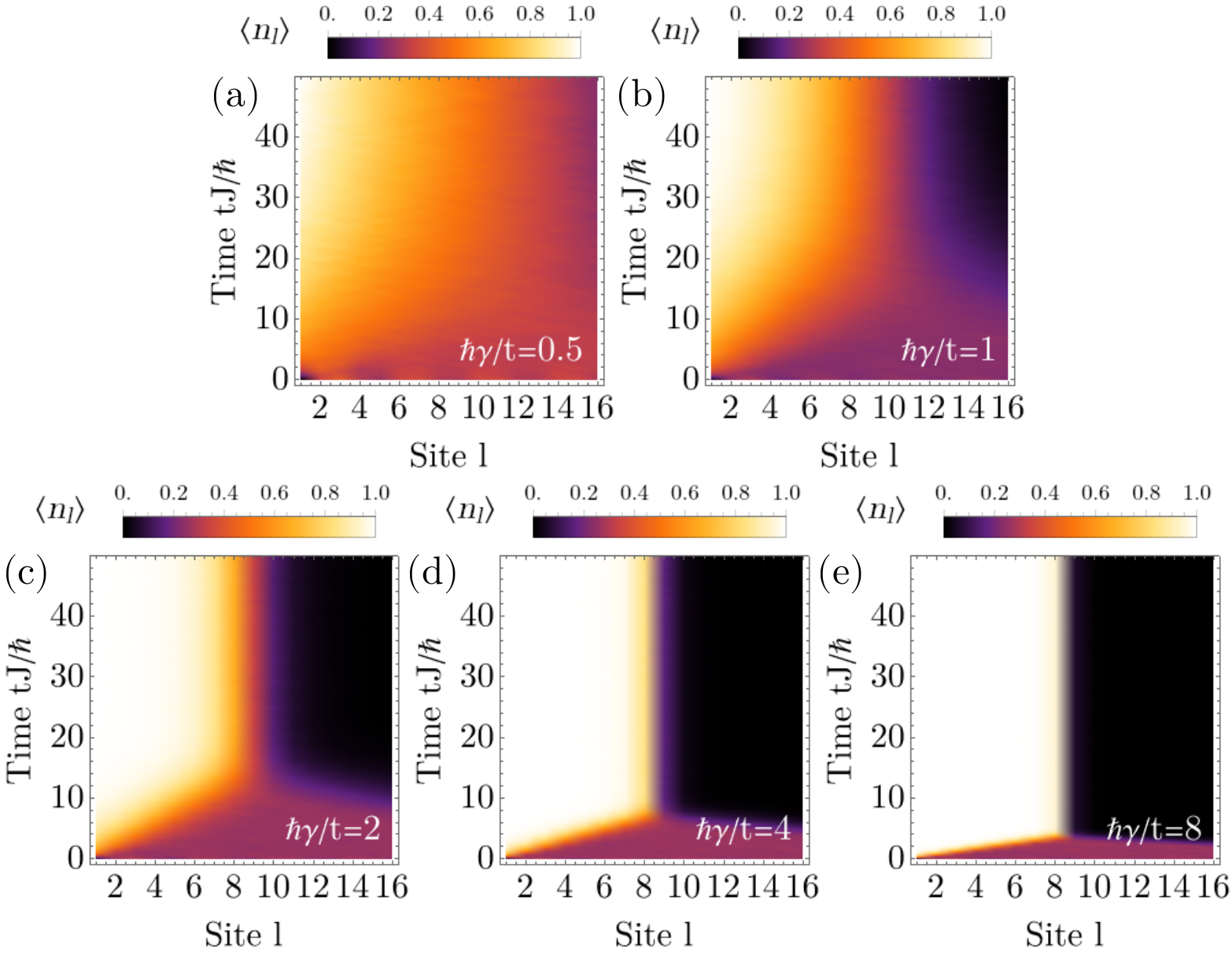}
\caption{The time dependence of the local densities, $\langle \hat{n}_l \rangle$, for the first half of the chain, $1\leq l \leq 16$, for the local atomic model with kinetic dissipation, Eqs.~(\ref{eq:Lindblad_atoms})-(\ref{eq:Hamiltonian_BH}). The different panels correspond to different values of the dissipation strength, $\hbar\gamma/t\in\{0.5,1,2,4,8\}$. We consider $L=32$ sites and $N=8$ particles.
}
\label{fig:density_time}
\end{figure}

\begin{figure}[!hbtp]
\centering
\includegraphics[width=0.4\textwidth]{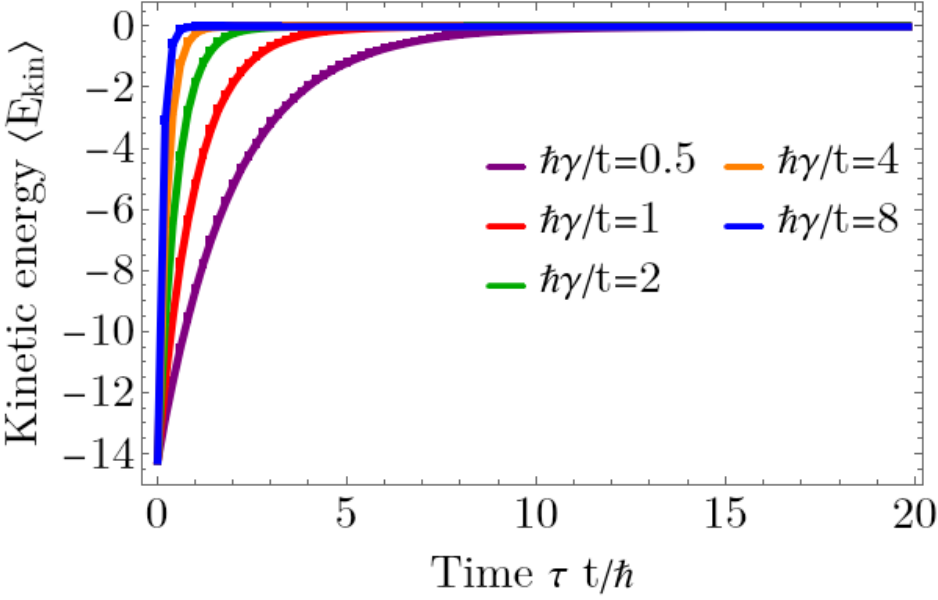}
\caption{The time dependence of the kinetic energy, $\hat{H}_{\text{kin}}$, for the local atomic model with kinetic dissipation, Eqs.~(\ref{eq:Lindblad_atoms})-(\ref{eq:Hamiltonian_BH}), and different values of the dissipation strength, $\hbar\gamma/t\in\{0.5,1,2,4,8\}$. 
We consider $L=32$ sites and $N=8$ particles.
}
\label{fig:kinetic_energy}
\end{figure}

\begin{figure}[!hbtp]
\centering
\includegraphics[width=0.48\textwidth]{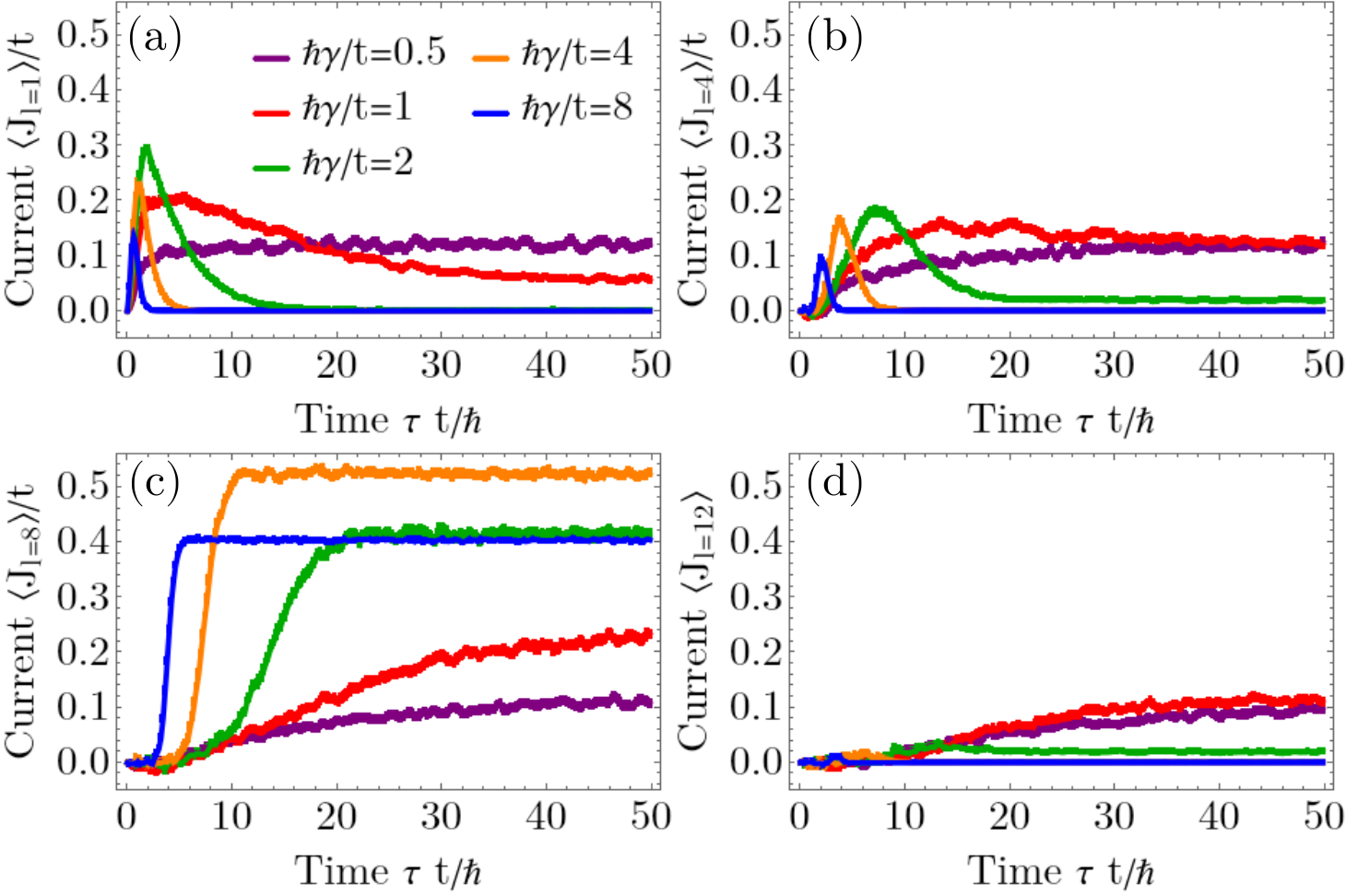}
\caption{The time dependence of the current, $\langle\hat{J}_l\rangle$, for the local atomic model with kinetic dissipation, Eqs.~(\ref{eq:Lindblad_atoms})-(\ref{eq:Hamiltonian_BH}), and different values of the dissipation strength, $\hbar\gamma/t\in\{0.5,1,2,4,8\}$. 
The different panels correspond to the different sites for which the $\langle\hat{J}_l\rangle$ was computed, $l\in\{1,4,8,12\}$.
We consider $L=32$ sites and $N=8$ particles.
}
\label{fig:current_kin_1}
\end{figure}

In the local atomic model, by turning on the nonreciprocal dissipation, Eq.~(\ref{eq:Lindblad_atoms}),  for a chain of bosonic atoms with hardcore interactions it is expected that the atoms will move towards the left end of the chain for the open boundary conditions considered.
We see this behavior in Fig.~\ref{fig:density}, where we plot the density profile at the time $\tau t/\hbar=50$ for different values of the dissipation strength. When dissipation dominates a state with a sharp interface between the filled and empty sites is the steady state, as obtained for the curve corresponding to $\hbar\gamma/t=8$. However, if we decrease the value of $\gamma$ the boundary between the filled and empty sites becomes smoother, as for $\hbar\gamma/t=2$ and $\hbar\gamma/t=4$.
The timescales for the density to stabilize to the shown profiles is inversely dependent on $\gamma$. We can observe in Fig.~\ref{fig:density_time}, where the dynamics of the local densities is depicted, that for large values of the dissipation strength the local densities quickly reach the steady state values.
In contrast, for $\hbar\gamma/t=0.5$ and $\hbar\gamma/t=1$ the steady state has not yet been reached for the considered final time [see Fig.~\ref{fig:density_time}(a) and Fig.~\ref{fig:density_time}(b)].
Thus, we expect that the density profiles shown in Fig.~\ref{fig:density} will become sharper around the sites $l=8$ and $l=9$ in the steady state.
In order to understand the dynamics of the atoms as the dissipation is turned on, it is interesting to analyze the behavior of the expectation value of the kinetic energy, $\langle\hat{H}_{\text{kin}}\rangle$, as depicted in Fig.~\ref{fig:kinetic_energy}.
We observe that the kinetic energy is quickly suppressed to zero for all values of the dissipation strength, with a time scale which is shorter than the timescales of the dynamics of the local densities (Fig.~\ref{fig:kinetic_energy} in comparison with Fig.~\ref{fig:density_time}).
In particular, the kinetic energy is zero even for $\hbar\gamma/t=0.5$ and $\hbar\gamma/t=1$, for which the local densities are still evolving at the latest times considered.
This implies that the coherence between neighboring sites, as measured by the real part of the single particle correlations $\Re\langle\hat{b}_{j}^\dagger \hat{b}_{j+1}\rangle$, i.e.~the tunneling terms, is rapidly decaying. In the following, we analyze the dynamical behavior of the imaginary part of the single particle correlations between neighboring sites, $\Im\langle\hat{b}_{j}^\dagger \hat{b}_{j+1}\rangle$, i.e.~the current terms given in Eq.~(\ref{eq:current_BH_av}), and of the dissipative contributions to the currents.

\begin{figure}[!hbtp]
\centering
\includegraphics[width=0.48\textwidth]{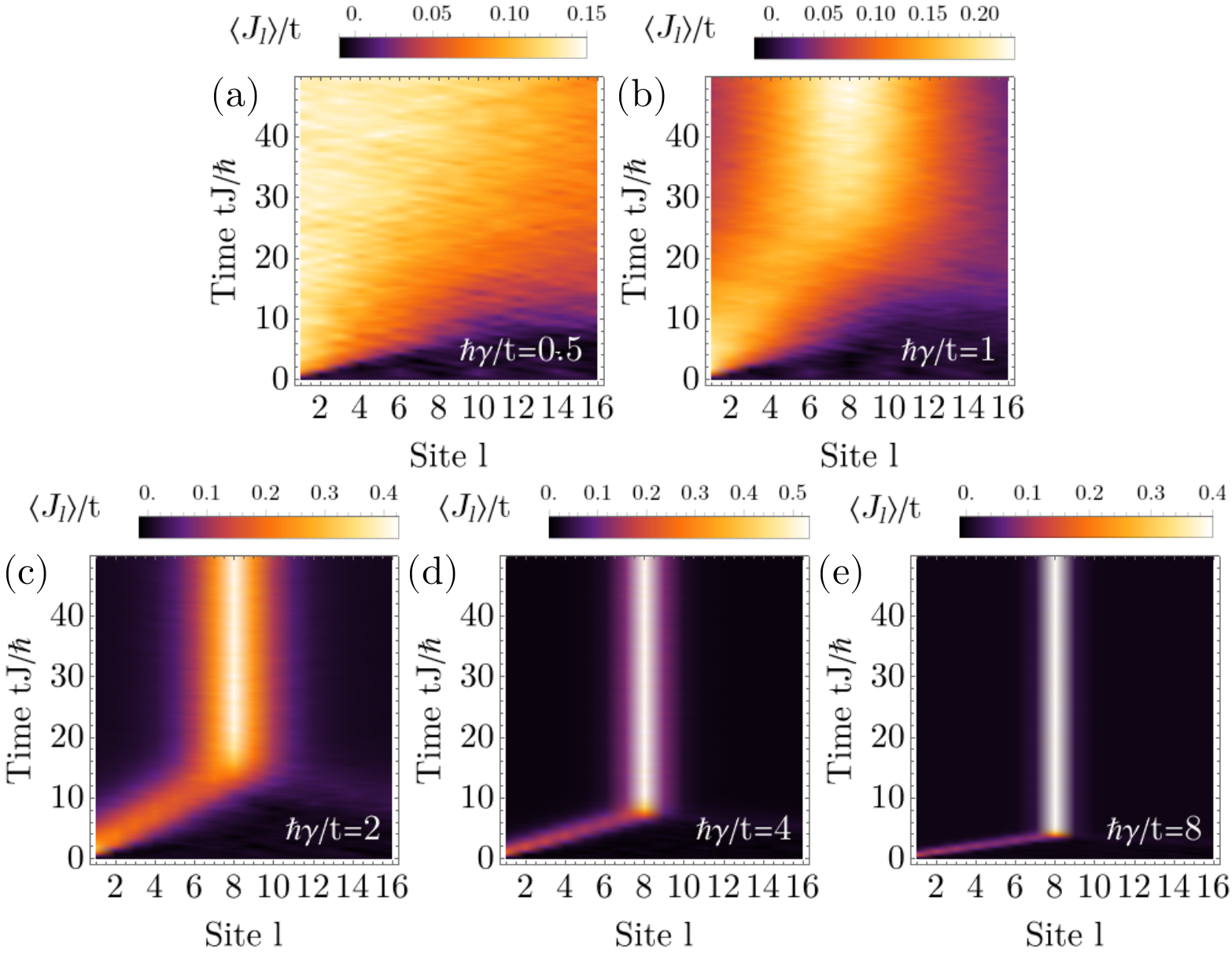}
\caption{The time dependence of the current, $\langle\hat{J}_l\rangle$, for the first half of the chain, $1\leq l \leq 16$, for the local atomic model with kinetic dissipation, Eqs.~(\ref{eq:Lindblad_atoms})-(\ref{eq:Hamiltonian_BH}). 
The different panels correspond to different values of the dissipation strength, $\hbar\gamma/t\in\{0.5,1,2,4,8\}$.
We consider $L=32$ sites and $N=8$ particles.
}
\label{fig:current_kin_2}
\end{figure}

\begin{figure}[!hbtp]
\centering
\includegraphics[width=0.4\textwidth]{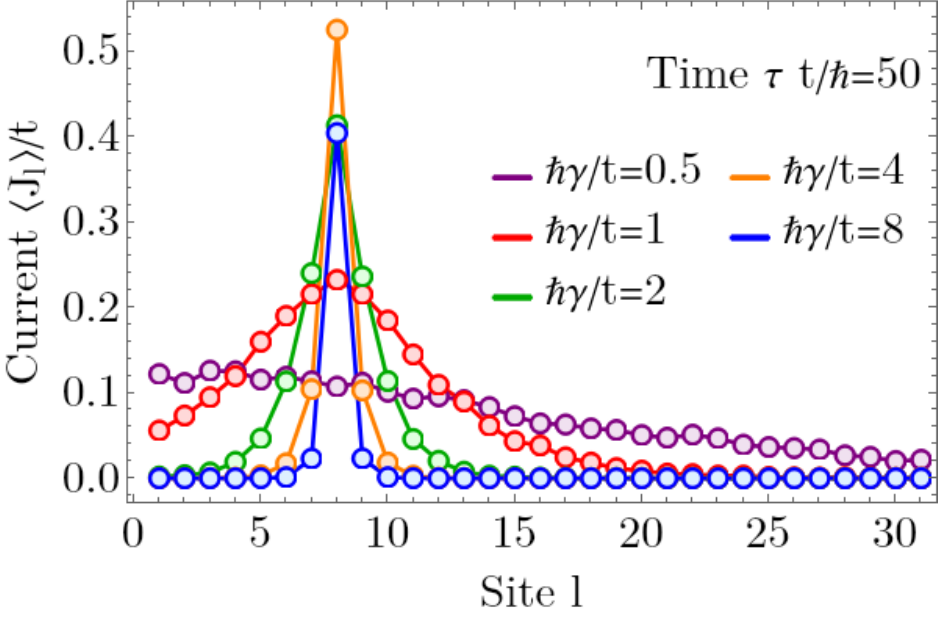}
\caption{The space dependence of the current, $\langle\hat{J}_l\rangle$ at time $\tau t/\hbar=50$, for the local atomic model with kinetic dissipation, Eqs.~(\ref{eq:Lindblad_atoms})-(\ref{eq:Hamiltonian_BH}), and different values of the dissipation strength, $\hbar\gamma/t\in\{0.5,1,2,4,8\}$. 
We consider $L=32$ sites and $N=8$ particles.
}
\label{fig:current_kin_3}
\end{figure}

\begin{figure}[!hbtp]
\centering
\includegraphics[width=0.48\textwidth]{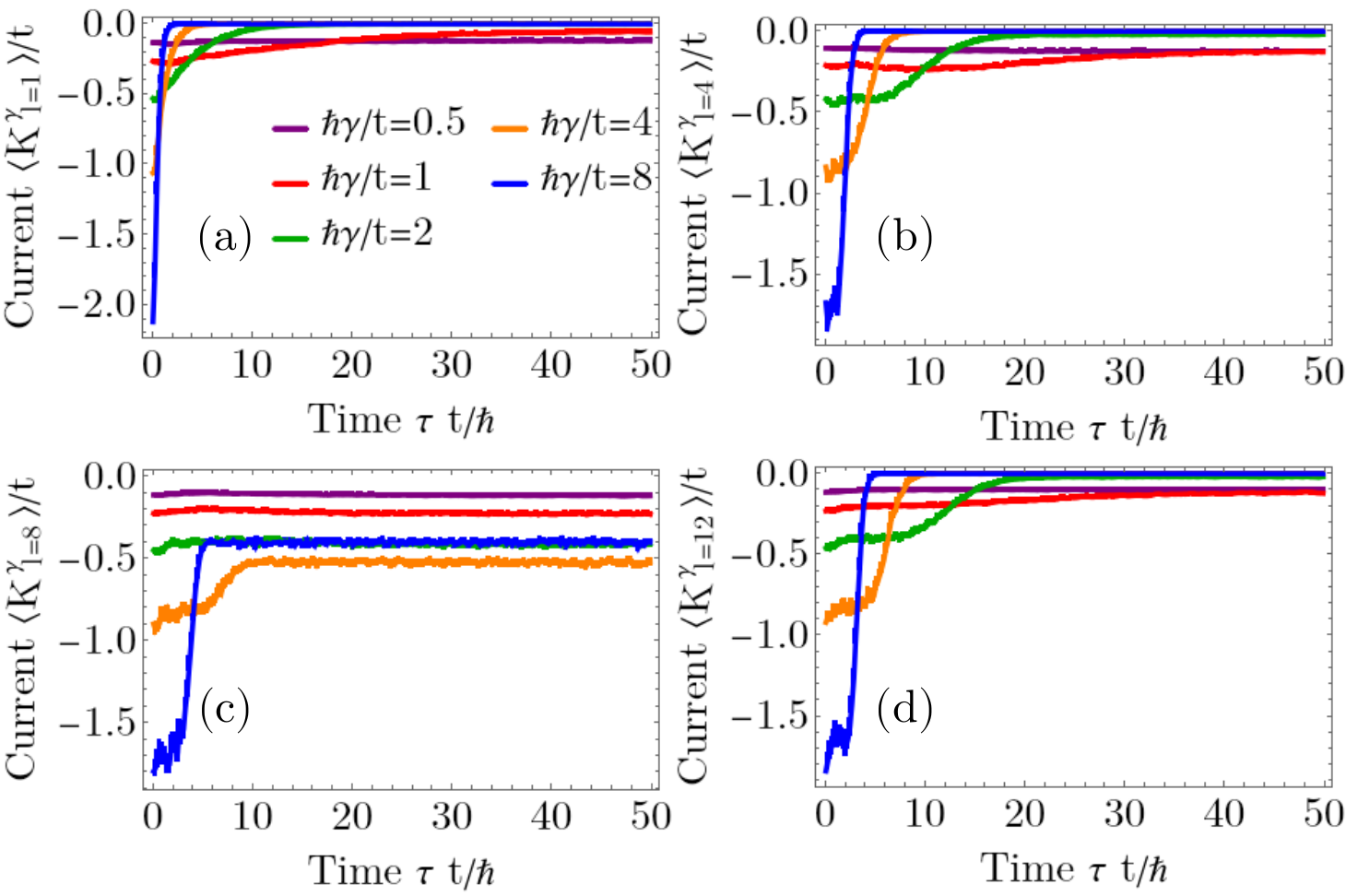}
\caption{The time dependence of the dissipative contribution to the current, $\langle\hat{K}^\gamma\rangle$, for the local atomic model with kinetic dissipation, Eqs.~(\ref{eq:Lindblad_atoms})-(\ref{eq:Hamiltonian_BH}), and different values of the dissipation strength, $\hbar\gamma/t\in\{0.5,1,2,4,8\}$. 
The different panels correspond to the different sites for which the $\langle\hat{K}^\gamma\rangle$ was computed, $l\in\{1,4,8,12\}$.
We consider $L=32$ sites and $N=8$ particles.
}
\label{fig:current_kin_gamma_1}
\end{figure}

In Eq.~(\ref{eq:current_kin_dis}) we saw that the dynamics of local densities for the local atomic model is controlled by currents stemming from the Hamiltonian tunneling terms, $\hat{J}_l$, and the dissipative coupling to kinetic terms, $\hat{K}^\gamma_l$. While in the steady state the local densities do not evolve anymore and their derivative is zero, this does not imply that the individual terms in Eq.~(\ref{eq:current_kin_dis}) are vanishing, thus, it is possible to get finite value of the currents originating from the Hamiltonian which are then exactly compensated by the currents having a dissipative origin.
We first explore the behavior of $\langle\hat{J}_l\rangle$, for which we show the dynamics in Fig.~\ref{fig:current_kin_1} and Fig.~\ref{fig:current_kin_2}.
We observe that for large dissipation strengths, $\hbar\gamma/t\geq 2$, for the first sites in the chain, $l\lesssim 6$, the current $\langle\hat{J}_l\rangle$ initially increases, signaling the redistribution of atomic density, and after reaching a maximum it is damped to a vanishing value. The time at which the maximum occurs in $\langle\hat{J}_l\rangle$ appears to have a linear dependence on the site $l$, with the slope increasing with $\hbar\gamma/t$. 
This behavior is observed in the plots corresponding to $\hbar\gamma/t\geq 2$ in Figs.~\ref{fig:current_kin_1}(a)-(b) and Figs.~\ref{fig:current_kin_2}(c)-(e).
For the sites that are empty at long times, $l\gtrsim 9$, $\langle\hat{J}_l\rangle$ has small values throughout the evolution [see Fig.~\ref{fig:current_kin_1}(d)]. 
In contrast, at the interface between the filled and empty sites for $l=8$, after an initial increase, the current $\langle\hat{J}_l\rangle$ stabilizes to a large finite value which persists in the steady state, as seen in Fig.~\ref{fig:current_kin_1}(c) and in Figs.~\ref{fig:current_kin_2}(c)-(e).
For the smaller values of the dissipation strength considered, $\hbar\gamma/t=0.5$ and $\hbar\gamma/t=1$, $\langle\hat{J}_l\rangle$ has not reached yet the steady state value. 
However, we observe that even though the maximal values of the currents $\langle\hat{J}_l\rangle$ are smaller compared to ones obtained for larger $\hbar\gamma/t$, there are finite currents also for sites away from $l=8$ at long times.
We see this also in Fig.~\ref{fig:current_kin_3}, where the space dependence of $\langle\hat{J}_l\rangle$ is shown at the time $\tau t/\hbar=50$. The width of the peak in $\langle\hat{J}_l\rangle$ occurring around $l=8$ becomes larger as the dissipation strength is lowered.
Interestingly, the largest value of the current $\langle\hat{J}_l\rangle$ for $l=8$ is obtained for $\hbar\gamma/t=4$, hinting that a finite optimal value of the dissipation strength might exist for obtaining a strong current.

\begin{figure}[!hbtp]
\centering
\includegraphics[width=0.48\textwidth]{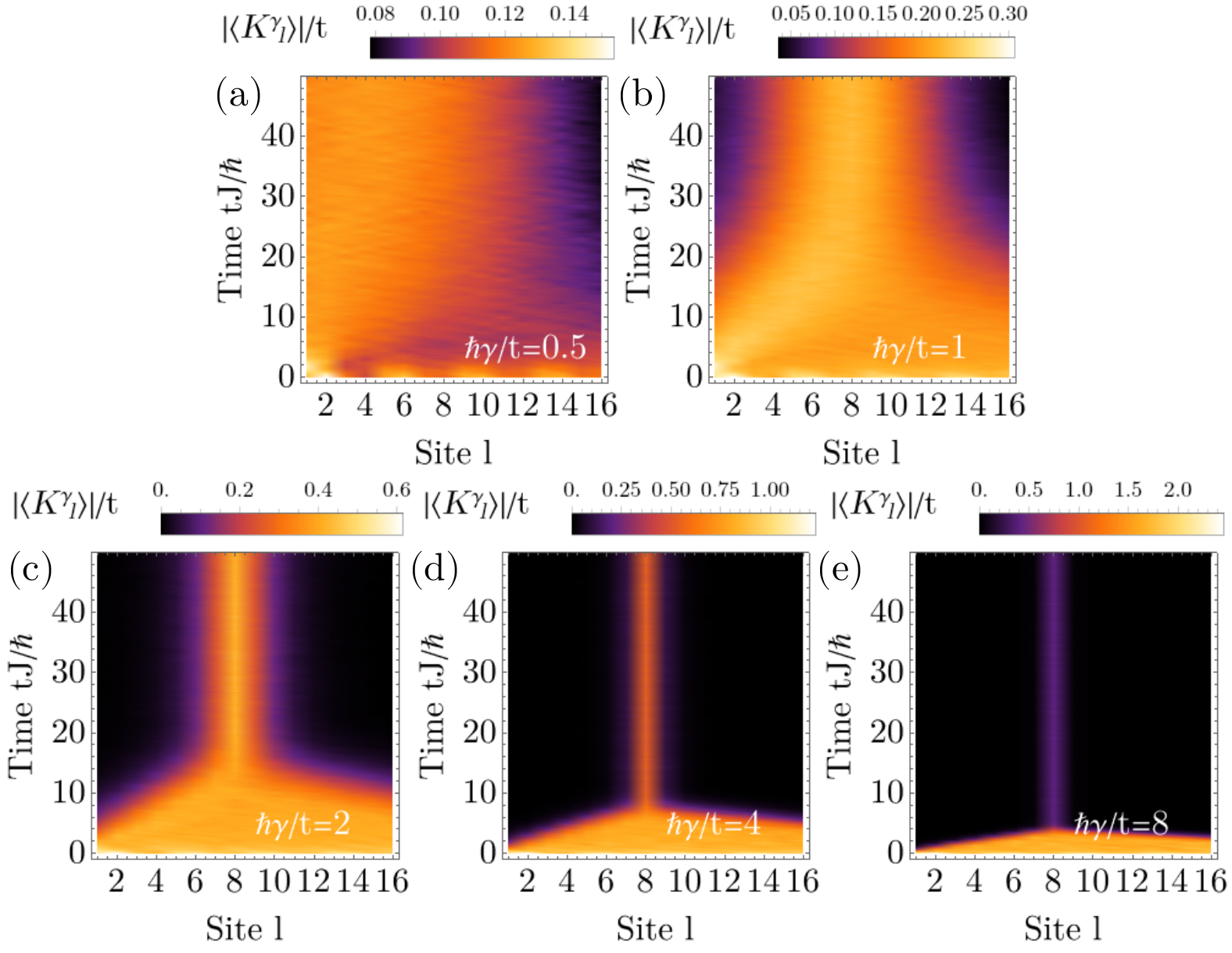}
\caption{The time dependence of the dissipative contribution to the current, $|\langle\hat{K}^\gamma\rangle|$, for the first half of the chain, $1\leq l \leq 16$, for the local atomic model with kinetic dissipation, Eqs.~(\ref{eq:Lindblad_atoms})-(\ref{eq:Hamiltonian_BH}). 
The different panels correspond to the different values of the dissipation strength, $\hbar\gamma/t\in\{0.5,1,2,4,8\}$.
We consider $L=32$ sites and $N=8$ particles.
}
\label{fig:current_kin_gamma_2}
\end{figure}

\begin{figure}[!hbtp]
\centering
\includegraphics[width=0.4\textwidth]{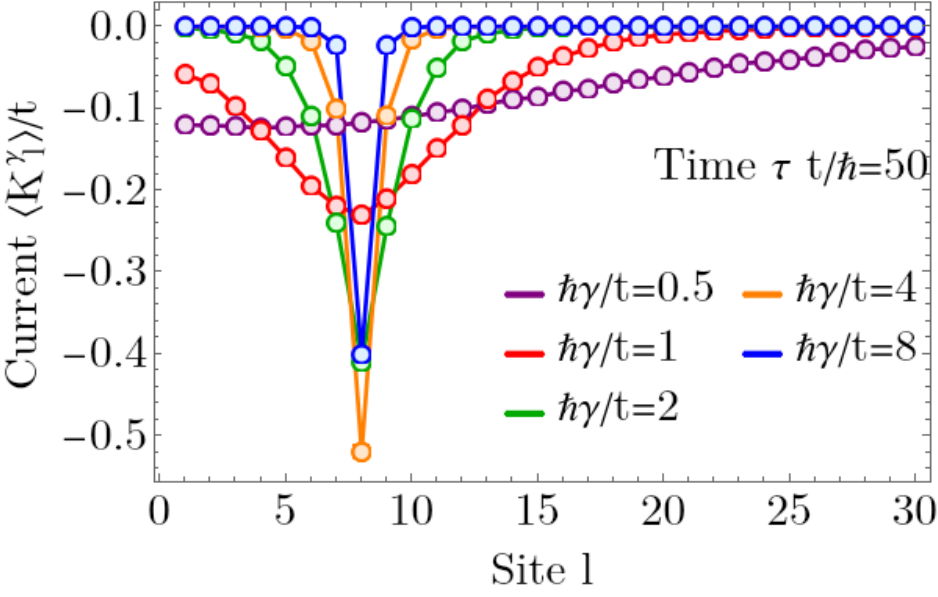}
\caption{The space dependence of the dissipative contribution to the current, $\langle\hat{K}^\gamma\rangle$ at time $\tau t/\hbar=50$, for the local atomic model with kinetic dissipation, Eqs.~(\ref{eq:Lindblad_atoms})-(\ref{eq:Hamiltonian_BH}), and different values of the dissipation strength, $\hbar\gamma/t\in\{0.5,1,2,4,8\}$. 
We consider $L=32$ sites and $N=8$ particles.
}
\label{fig:current_kin_gamma_3}
\end{figure}

\begin{figure}[!hbtp]
\centering
\includegraphics[width=0.4\textwidth]{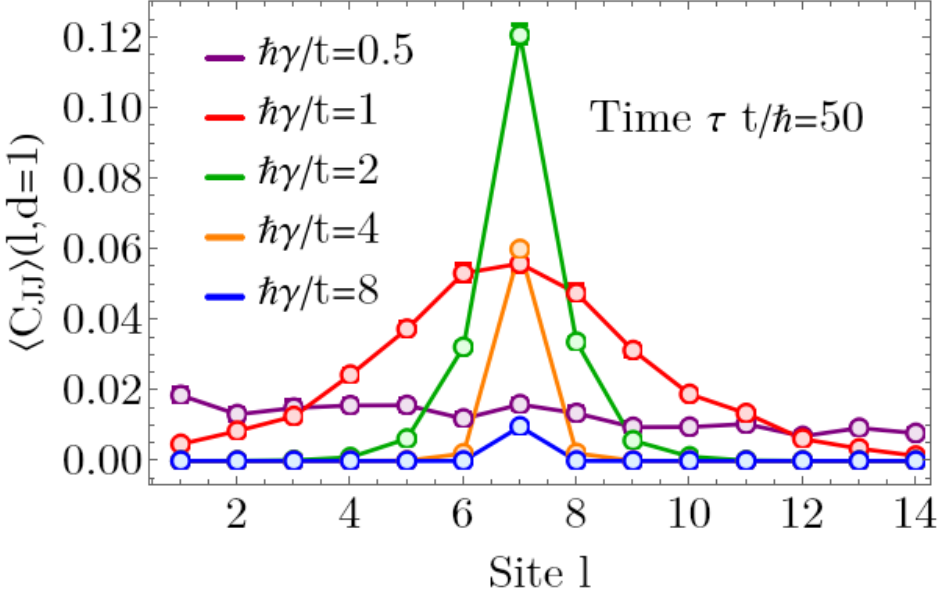}
\caption{The space dependence of the current-current correlations, $\langle\hat{C}_{JJ}\rangle(l,d)$ at time $\tau t/\hbar=50$ and distance $d=1$, for the local atomic model with kinetic dissipation, Eqs.~(\ref{eq:Lindblad_atoms})-(\ref{eq:Hamiltonian_BH}), and different values of the dissipation strength, $\hbar\gamma/t\in\{0.5,1,2,4,8\}$. 
We consider $L=32$ sites and $N=8$ particles.
}
\label{fig:current_corr_kin_1}
\end{figure}

We discuss in the following the dissipative contribution to the currents, $\langle\hat{K}^\gamma\rangle$, for the local atomic model, for which the time dependence is depicted in Fig.~\ref{fig:current_kin_gamma_1} and Fig.~\ref{fig:current_kin_gamma_2}. As shown in Eq.~(\ref{eq:current_kin_dis_av}) the current $\langle\hat{K}^\gamma\rangle$ is given by the expectation values of density-density correlations and local densities, thus, as these observables have finite values in the initial state $\langle\hat{K}^\gamma\rangle$ is non-zero as soon as we turn on the dissipative processes.
We observe in Fig.~\ref{fig:current_kin_gamma_1} that for large dissipation strengths, $\hbar\gamma/t\geq 2$, $\langle\hat{K}^\gamma\rangle$ has a large value at short times, corresponding to the rapid dynamics observed in the local densities in Fig.~\ref{fig:density_time}, which afterwards relaxes to the same spatial profile as obtained for $\langle\hat{J}_l\rangle$ (Fig.~\ref{fig:current_kin_gamma_3} in comparison with Fig.~\ref{fig:current_kin_3}).
Thus, after the short time dynamics $\langle\hat{K}^\gamma\rangle$ acquires a peak at $l=8$ with it width dependent on the value of $\hbar\gamma/t$, as seen in Figs.~\ref{fig:current_kin_gamma_2}(c)-(e).
For the smaller values of the dissipation strength, $\hbar\gamma/t=0.5$ and $\hbar\gamma/t=1$, the values of $\langle\hat{K}^\gamma\rangle$ are initially smaller than in the case of the stronger dissipation, however, they remain finite at the longest times considered throughout the chain [see Fig.~\ref{fig:current_kin_gamma_1} and Figs.~\ref{fig:current_kin_gamma_2}(a)-(b)] determining the dynamics of the local densities.

\begin{figure}[!hbtp]
\centering
\includegraphics[width=0.48\textwidth]{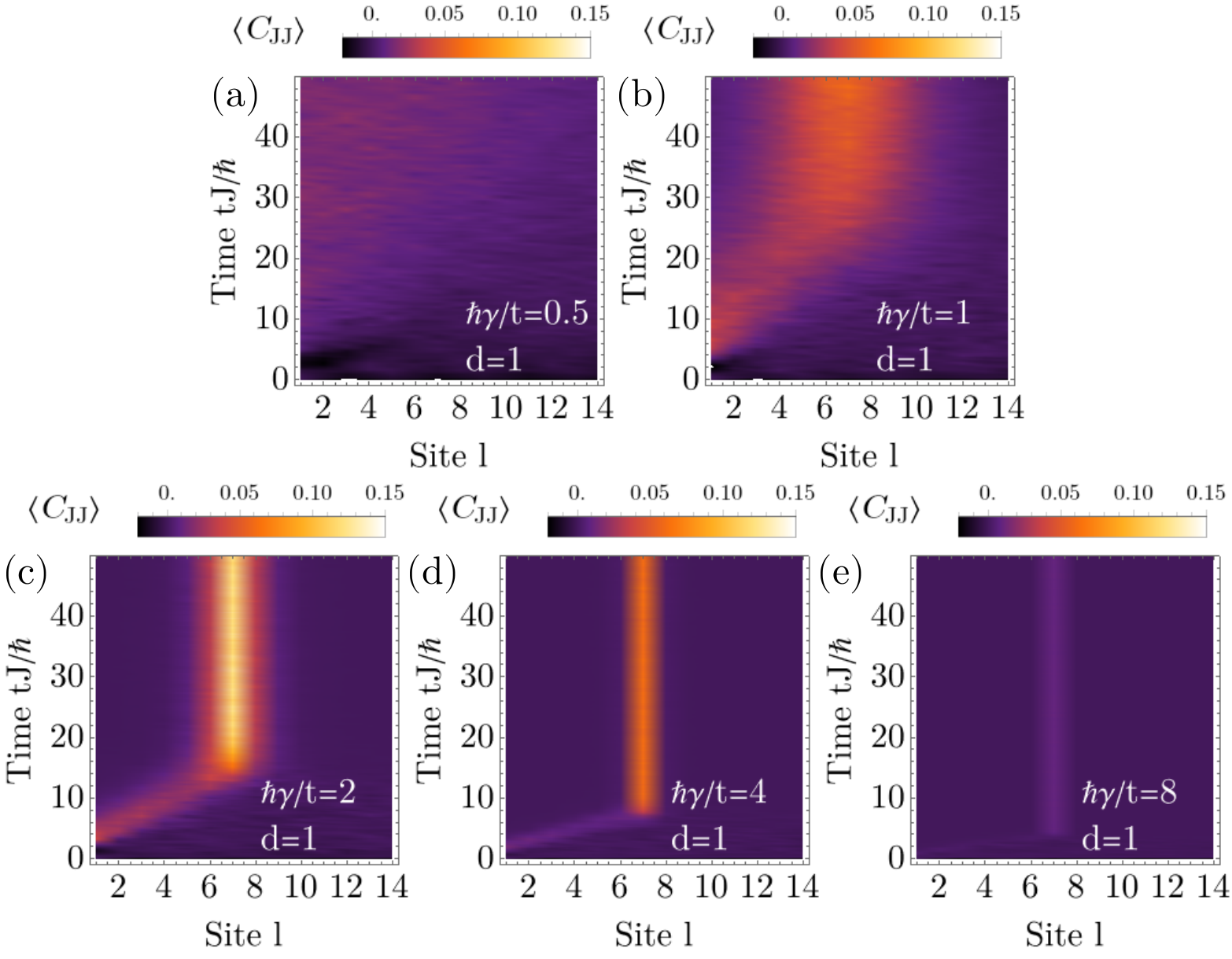}
\caption{The time dependence of the current-current correlations, $\langle\hat{C}_{JJ}\rangle(l,d)$, for distance $d=1$ and the sites $1\leq l \leq 14$, for the local atomic model with kinetic dissipation, Eqs.~(\ref{eq:Lindblad_atoms})-(\ref{eq:Hamiltonian_BH}).
The different panels correspond to the different values of the dissipation strength, $\hbar\gamma/t\in\{0.5,1,2,4,8\}$.
We consider $L=32$ sites and $N=8$ particles.
}
\label{fig:current_corr_kin_2}
\end{figure}

\begin{figure}[!hbtp]
\centering
\includegraphics[width=0.48\textwidth]{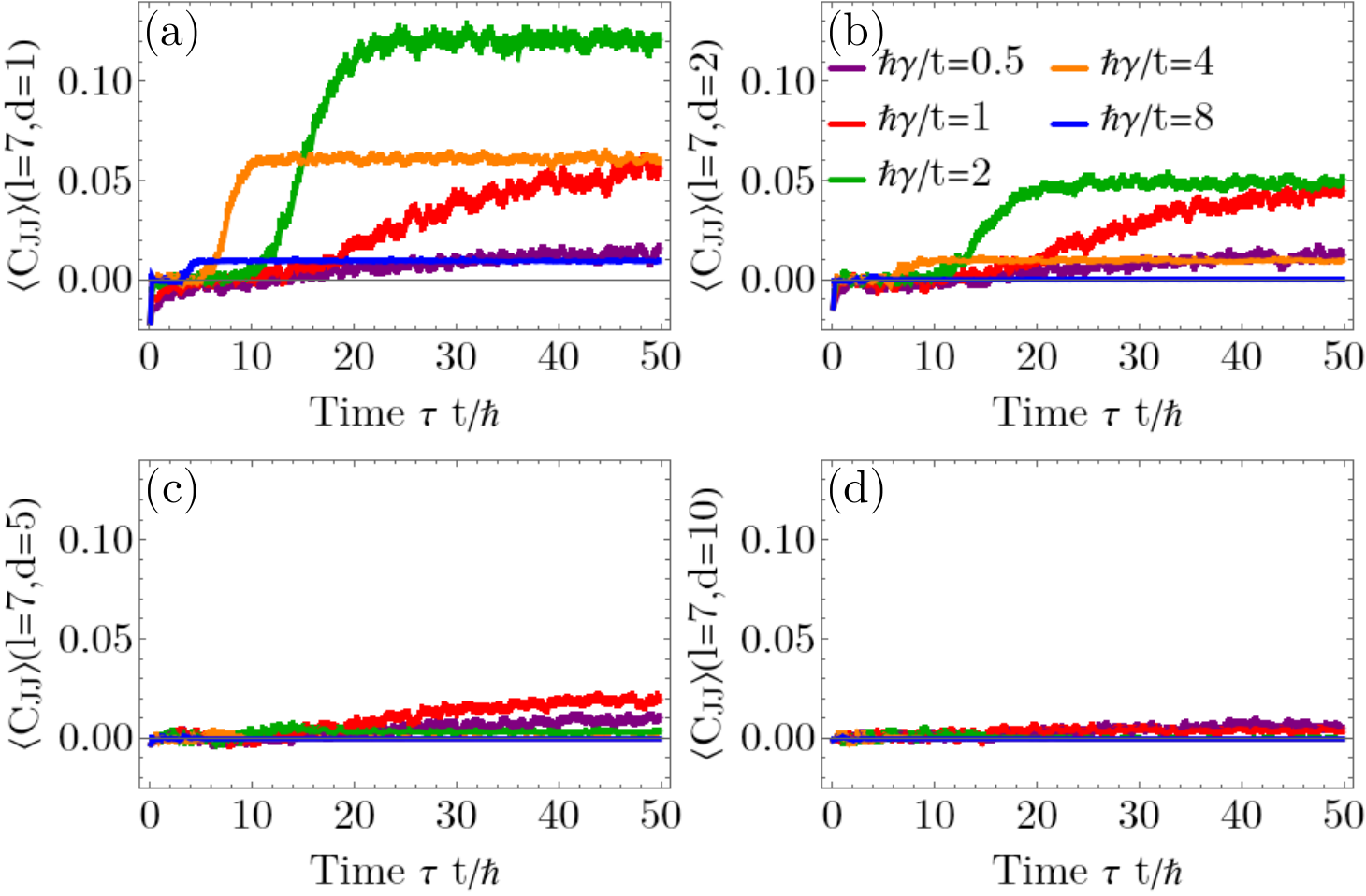}
\caption{The time dependence of the current-current correlations, $\langle\hat{C}_{JJ}\rangle(l,d)$ for the site $l=7$, for the local atomic model with kinetic dissipation, Eqs.~(\ref{eq:Lindblad_atoms})-(\ref{eq:Hamiltonian_BH}), and different values of the dissipation strength, $\hbar\gamma/t\in\{0.5,1,2,4,8\}$. 
The different panels correspond to the different distances for which the $\hat{C}_{JJ}(l,d)$ was computed, $d\in\{1,2,5,10\}$.
We consider $L=32$ sites and $N=8$ particles.
}
\label{fig:current_corr_kin_3}
\end{figure}

\begin{figure}[!hbtp]
\centering
\includegraphics[width=0.4\textwidth]{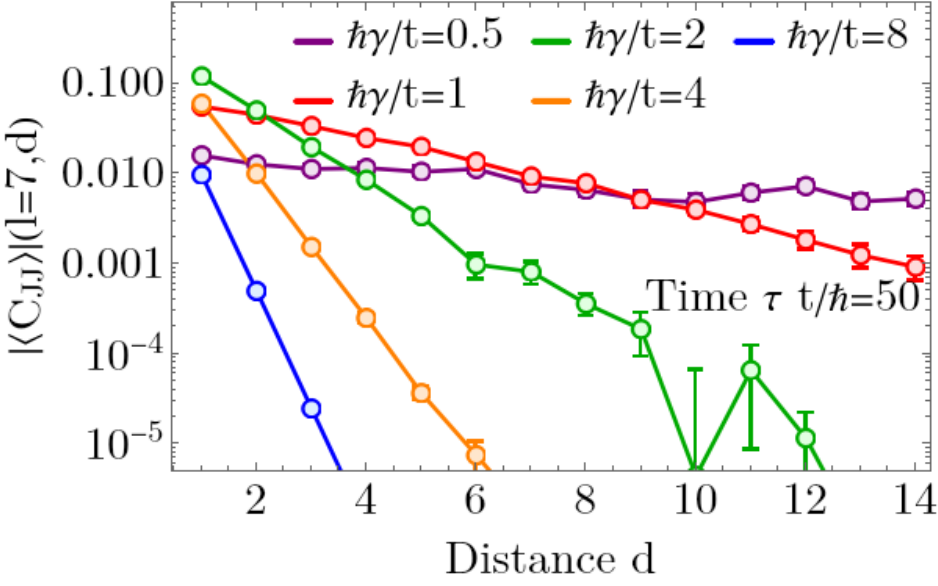}
\caption{The distance dependence of the current-current correlations, $|\langle\hat{C}_{JJ}\rangle|(l,d)$ at time $\tau t/\hbar=50$ and the site $l=7$, for the local atomic model with kinetic dissipation, Eqs.~(\ref{eq:Lindblad_atoms})-(\ref{eq:Hamiltonian_BH}), and different values of the dissipation strength, $\hbar\gamma/t\in\{0.5,1,2,4,8\}$. 
We use a semi-log scale for the decay of the correlations.
We consider $L=32$ sites and $N=8$ particles.
}
\label{fig:current_corr_kin_4}
\end{figure}

In the final part of this section, we discuss the behavior of the current-current correlations $\hat{C}_{JJ}(l,d)$, defined in Eq.~(\ref{eq:current_current}), for the local atomic model. These correlations measure the coherence between the currents which arise from the Hamiltonian tunneling terms.
We aim to identify if such correlations can have a finite value and how they are behaving if the distance $d$ between the sites which are probed increases.
In Fig.~\ref{fig:current_corr_kin_1} we plot the short distance, $d=1$, behavior of $\hat{C}_{JJ}$ throughout the chain at the final time considered, while the dynamics is shown in Fig.~\ref{fig:current_corr_kin_2}.
Similar to the behavior of the local currents $\langle\hat{J}_l\rangle$, shown in Fig.~\ref{fig:current_kin_3}, the largest value is obtained for the sites involving the interface between the filled and empty sites, for $\hat{C}_{JJ}(l=7,d=1)$. As shown in Figs.~\ref{fig:current_corr_kin_2}(b)-(e) the maximum of $\hat{C}_{JJ}(l,d=1)$ moves from the first sites towards $l=7$ were it stabilizes at long times, with the speed at short times and the width of the spatial profile throughout the evolution dependent on the strength of $\gamma$.
Interestingly, the largest value of $\hat{C}_{JJ}(l,d=1)$ as a function of the dissipation strength is for the intermediate value of $\hbar\gamma/t=2$ [see Fig.~\ref{fig:current_corr_kin_1} and Fig.~\ref{fig:current_corr_kin_2}(c)]. In particular, $\hbar\gamma/t=8$ we obtain very small values of the correlations [see Fig.~\ref{fig:current_corr_kin_1} and Fig.~\ref{fig:current_corr_kin_2}(e)], which we attribute to the very narrow spatial distribution of $\langle\hat{J}_l\rangle$ for this value of dissipation (see Fig.~\ref{fig:current_kin_3}) and due to the loss of coherence for large dissipation strengths.

To analyze the behavior with the distance $d$ of the current-current correlations $\hat{C}_{JJ}(l,d)$ we focus on the starting site on which we obtain the largest values for $d=1$, namely $l=7$.
We plot the time dependence of $\hat{C}_{JJ}(l=7,d)$ for several distances in Fig.~\ref{fig:current_corr_kin_3} and the correlations as a function of $d$ for $\tau t/\hbar=50$ in Fig.~\ref{fig:current_corr_kin_4}. 
We obtain that while finite values of the correlations are obtained at short distances for $1\leq\hbar\gamma/t\leq 4$, $\hat{C}_{JJ}(l,d)$ is rapidly suppressed with the distance. In particular, within the accuracy of our numerical data for the correlations at longer distances, the decay of the current-current correlations, $\hat{C}_{JJ}$, seems to be exponential with the distance $d$ for $\hbar\gamma/t\geq 1$, with a faster decay for stronger dissipation strengths.

To summarize the results of this section, we obtained that for a chain of hardcore bosons, on which dissipation acts through kinetic terms, local currents can arise when dissipation dominates the dynamics. However, the finite values of the currents at long times are confined to the sites around the interface between the filled and empty sites. The currents stemming from the Hamiltonian tunneling terms are compensated in the steady state by the dissipative current contributions.
Finite current-current correlations are obtained for these sites, albeit with of small value and quickly decaying with the distance.
In the following section, we investigate how the same observables behave with the atoms experience the non-reciprocal dissipative processes mediated through the field of an optical cavity.

\subsection{Dynamics of hardcore bosons coupled to a dissipative cavity \label{sec:results_cavity}}

\begin{figure}[!hbtp]
\centering
\includegraphics[width=0.4\textwidth]{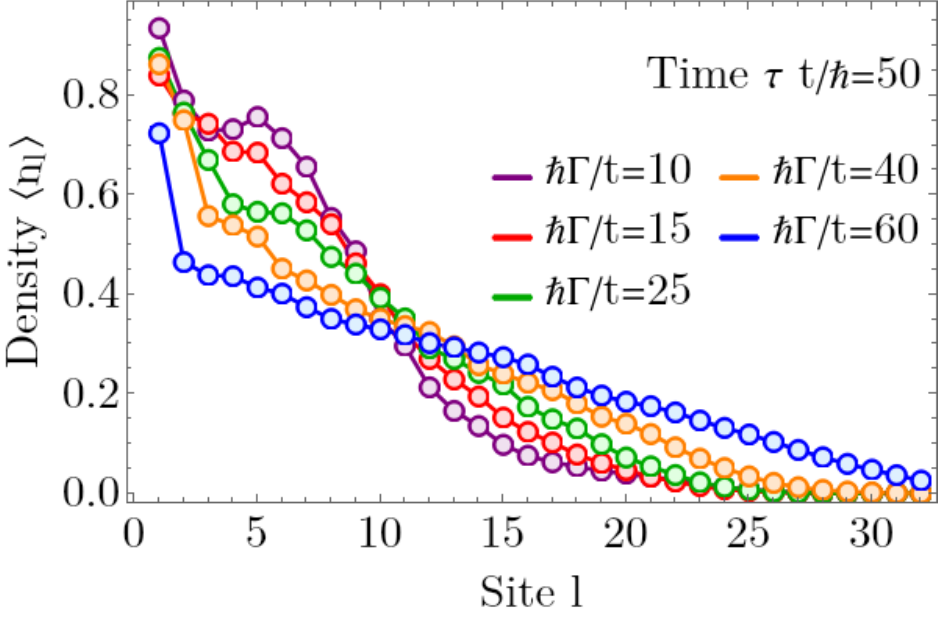}
\caption{The local density profile, $\langle \hat{n}_l \rangle$, at time $\tau t/\hbar=50$, for the global atom-cavity model, Eqs.~(\ref{eq:Lindblad})-(\ref{eq:Hamiltonian_ac}), and different values of the dissipation strength, $\hbar\Gamma/t\in\{10,15,25,40,60\}$. The other parameters are $L=32$ sites, $N=8$ particles, detuning $\hbar\delta/t=5$ and atoms-cavity coupling strength $\hbar\Omega/t=2.5$.
}
\label{fig:density_cav}
\end{figure}

\begin{figure}[!hbtp]
\centering
\includegraphics[width=0.48\textwidth]{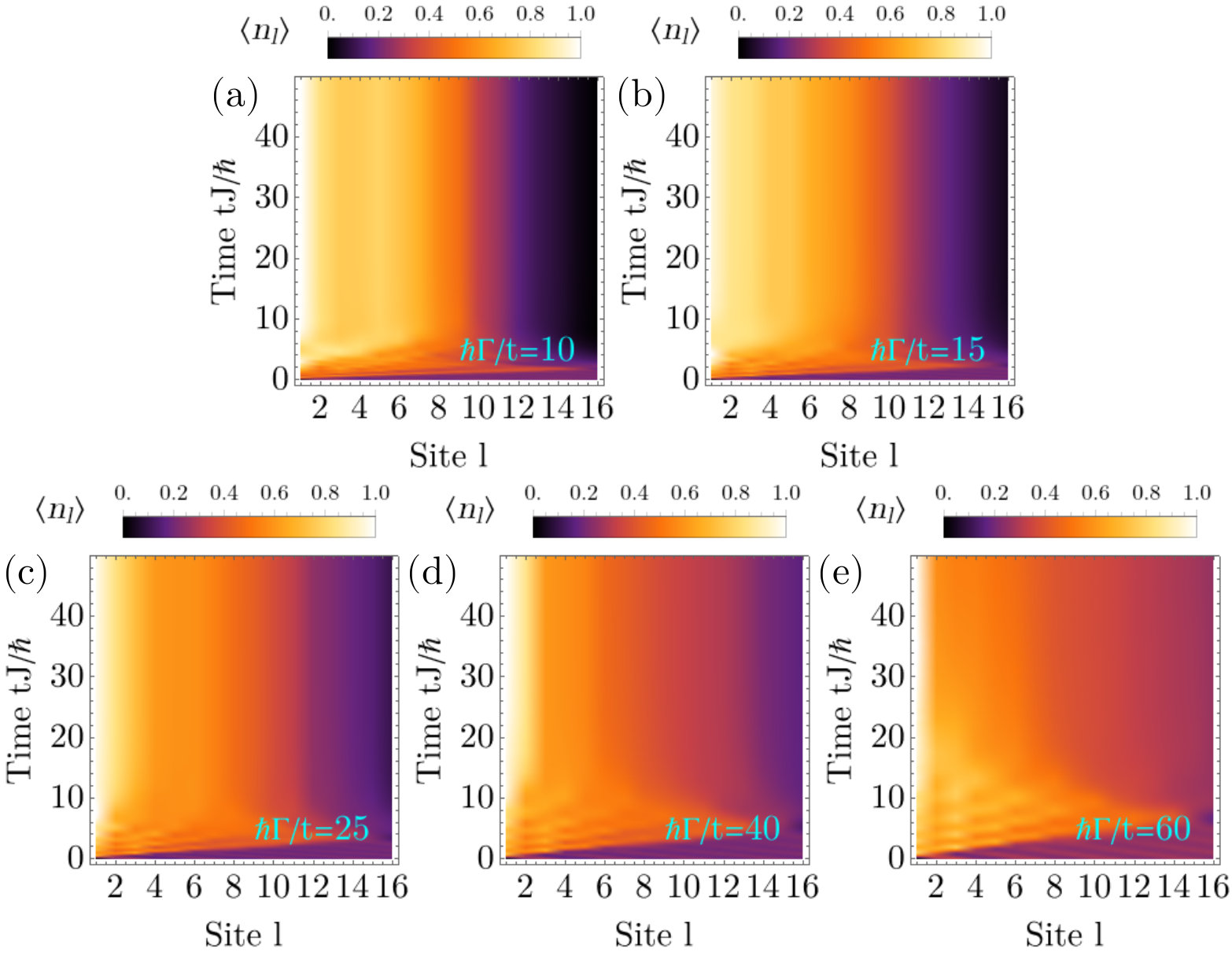}
\caption{The time dependence of the local densities, $\langle \hat{n}_l \rangle$, for the first half of the chain, $1\leq l \leq 16$, for the global atom-cavity model, Eqs.~(\ref{eq:Lindblad})-(\ref{eq:Hamiltonian_ac}). The different panels correspond to different values of the dissipation strength, $\hbar\Gamma/t\in\{10,15,25,40,60\}$. The other parameters are $L=32$ sites, $N=8$ particles, detuning $\hbar\delta/t=5$ and atoms-cavity coupling strength $\hbar\Omega/t=2.5$.
}
\label{fig:density_time_cav}
\end{figure}

\begin{figure}[!hbtp]
\centering
\includegraphics[width=0.4\textwidth]{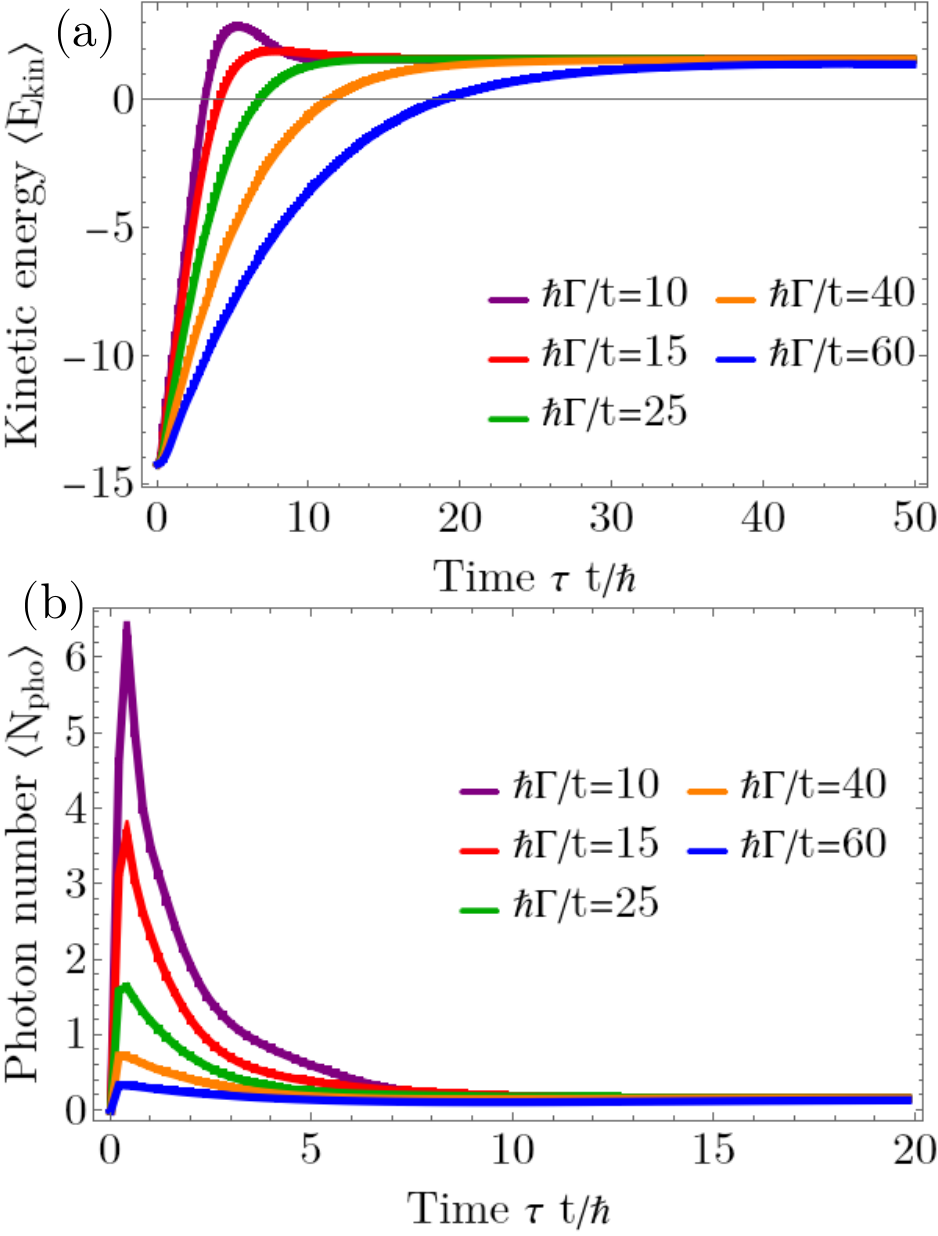}
\caption{(a) The time dependence of the kinetic energy, $\hat{H}_{\text{kin}}$, for the global atom-cavity model, Eqs.~(\ref{eq:Lindblad})-(\ref{eq:Hamiltonian_ac}). 
(b) The time dependence of the photon number, $N_\text{pho}=a^\dagger a$.
We show different values of the dissipation strength, $\hbar\Gamma/t\in\{10,15,25,40,60\}$. The other parameters are $L=32$ sites, $N=8$ particles, detuning $\hbar\delta/t=5$ and atoms-cavity coupling strength $\hbar\Omega/t=2.5$.
}
\label{fig:kinetic_energy_photon_no}
\end{figure}

\begin{figure}[!hbtp]
\centering
\includegraphics[width=0.48\textwidth]{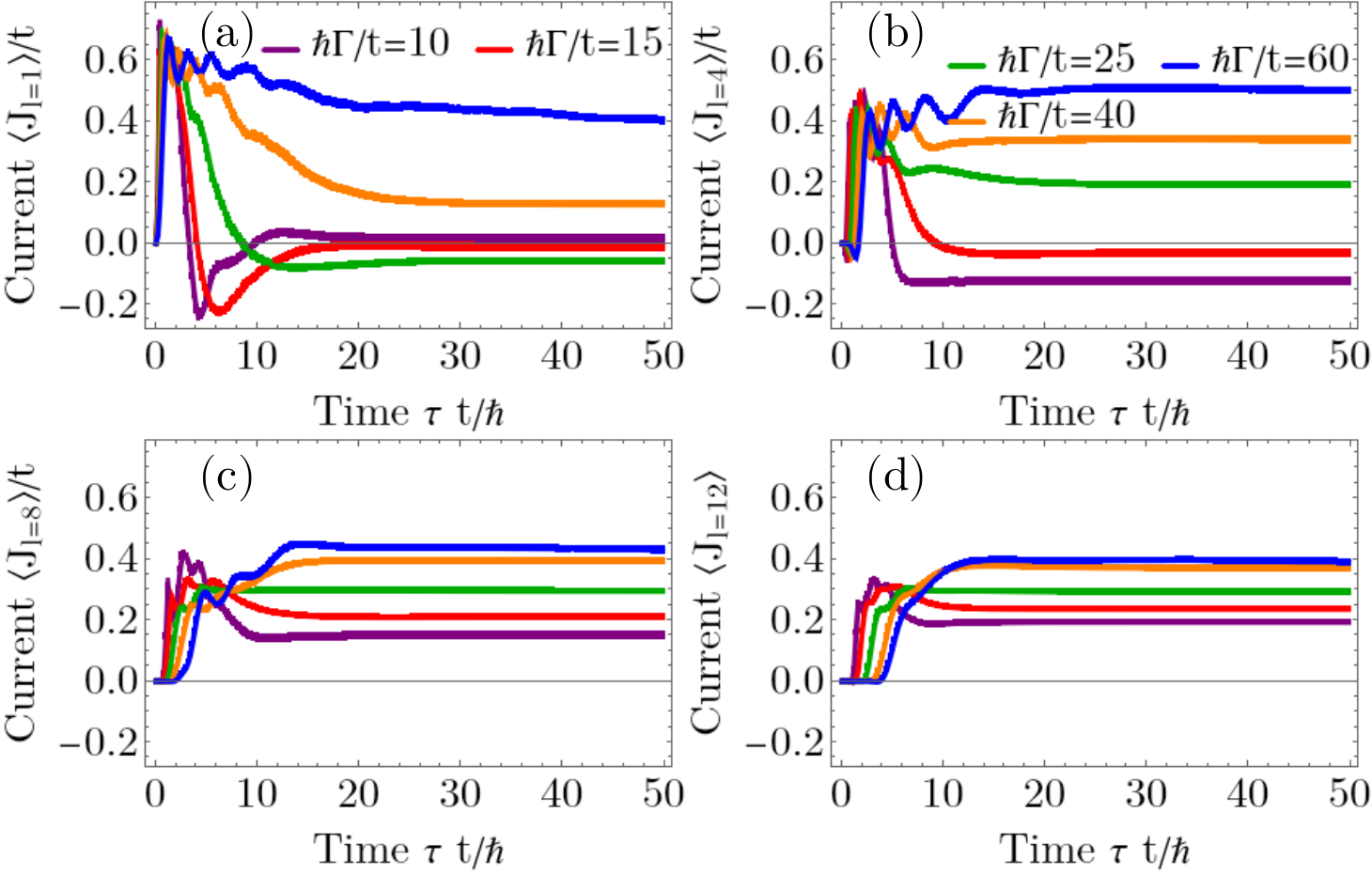}
\caption{The time dependence of the current, $\langle\hat{J}_l\rangle$, for the global atom-cavity model, Eqs.~(\ref{eq:Lindblad})-(\ref{eq:Hamiltonian_ac}), and different values of the dissipation strength, $\hbar\Gamma/t\in\{10,15,25,40,60\}$. 
The different panels correspond to the different sites for which the $\langle\hat{J}_l\rangle$ was computed, $l\in\{1,4,8,12\}$.
The other parameters are $L=32$ sites, $N=8$ particles, detuning $\hbar\delta/t=5$ and atoms-cavity coupling strength $\hbar\Omega/t=2.5$.
}
\label{fig:current_cav_1}
\end{figure}

In this section, we analyze the results of the atom-cavity model, Eqs.~(\ref{eq:Lindblad})-(\ref{eq:Hamiltonian_ac}), in which the cavity is globally coupled to atomic tunneling terms.
As the cavity is under the action of photon losses this is a nonreciprocal scenario, in which one of the atomic tunneling directions is favored  compared to the other.
We can see this behavior in Fig.~\ref{fig:density_cav}, where the density profile at time $\tau t/\hbar=50$, and in Fig.~\ref{fig:density_time_cav}, where the time evolution of the local densities, are shown.
The atoms accumulate towards the beginning of the chain, however, there is not a sharp interface between filled and empty sites as obtained for the atomic under kinetic dissipation model (see Fig.~\ref{fig:density}).
This can be understood from the fact that for the effective atomic dissipative channel in the presence of the coupling to the cavity, controlled by $\mathcal{O}=\sum_j  \hat{b}_{j}^\dagger \hat{b}_{j+1}$, the decoherence free subspace contains other states beside $\ket{1\dots 10\dots0}$.
Furthermore, in the strongly dissipative regime on which we have focused, we observe that by increasing the photonic dissipation strength $\Gamma$, we obtain a smoother crossover between the low density and high density sites, see Fig.~\ref{fig:density_cav}.
As the effective atomic dissipation strength is $\propto 1/\Gamma$, we also obtained a slow down of the dynamics as $\Gamma$ is increased, see Fig.~\ref{fig:density_time_cav}, due to the Zeno effect (this can also be identified in the dynamics of the some of the other observables presented in the following).
For all the dissipation strengths considered it seems that at the final evolution time we are either in, or very close, to the steady state of the coupled atom-cavity system.

In Fig.~\ref{fig:kinetic_energy_photon_no}(b) we depict the dynamics of the photon number for different values of the dissipation strength $\Gamma$. As the initial state is the ground state of the hardcore bosonic chain which contains a finite expectation value for the tunneling terms to which the cavity is coupled, at short times we have a rapid increase in the number of photons. The height of the initial peak is larger for the smaller values of $\Gamma$. After the initial increase, the photon number decays to a small, but finite value, at which it stabilizes towards the steady state.

\begin{figure}[!hbtp]
\centering
\includegraphics[width=0.48\textwidth]{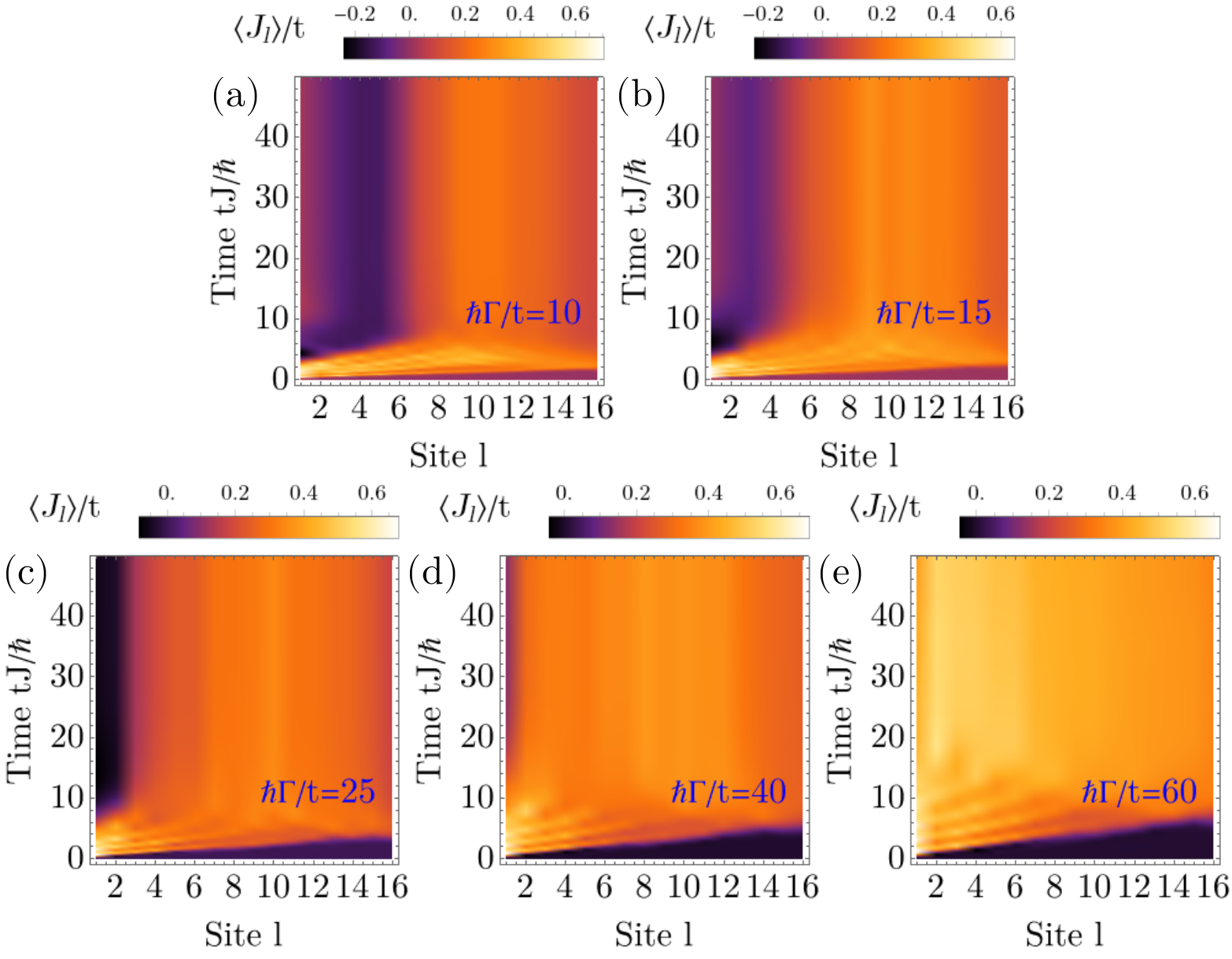}
\caption{The time dependence of the current, $\langle\hat{J}_l\rangle$, for the first half of the chain, $1\leq l \leq 16$, for the global atom-cavity model, Eqs.~(\ref{eq:Lindblad})-(\ref{eq:Hamiltonian_ac}). 
The different panels correspond to different values of the dissipation strength, $\hbar\Gamma/t\in\{10,15,25,40,60\}$.
The other parameters are $L=32$ sites, $N=8$ particles, detuning $\hbar\delta/t=5$ and atoms-cavity coupling strength $\hbar\Omega/t=2.5$.
}
\label{fig:current_cav_2}
\end{figure}

\begin{figure}[!hbtp]
\centering
\includegraphics[width=0.4\textwidth]{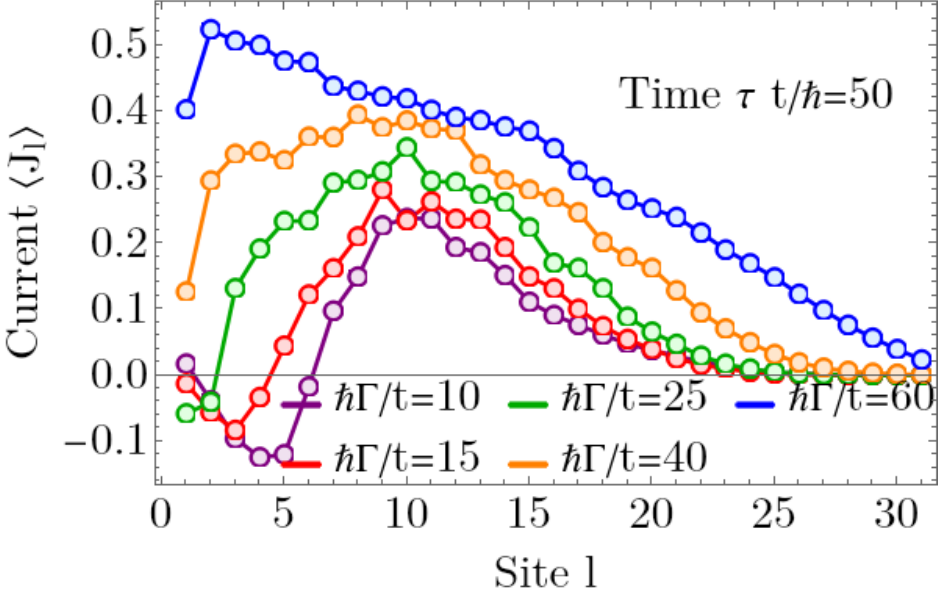}
\caption{The space dependence of the current, $\langle\hat{J}_l\rangle$ at time $\tau t/\hbar=50$, for the global atom-cavity model, Eqs.~(\ref{eq:Lindblad})-(\ref{eq:Hamiltonian_ac}), and different values of the dissipation strength, $\hbar\Gamma/t\in\{10,15,25,40,60\}$. 
The other parameters are $L=32$ sites, $N=8$ particles, detuning $\hbar\delta/t=5$ and atoms-cavity coupling strength $\hbar\Omega/t=2.5$.
}
\label{fig:current_cav_3}
\end{figure}

\begin{figure}[!hbtp]
\centering
\includegraphics[width=0.48\textwidth]{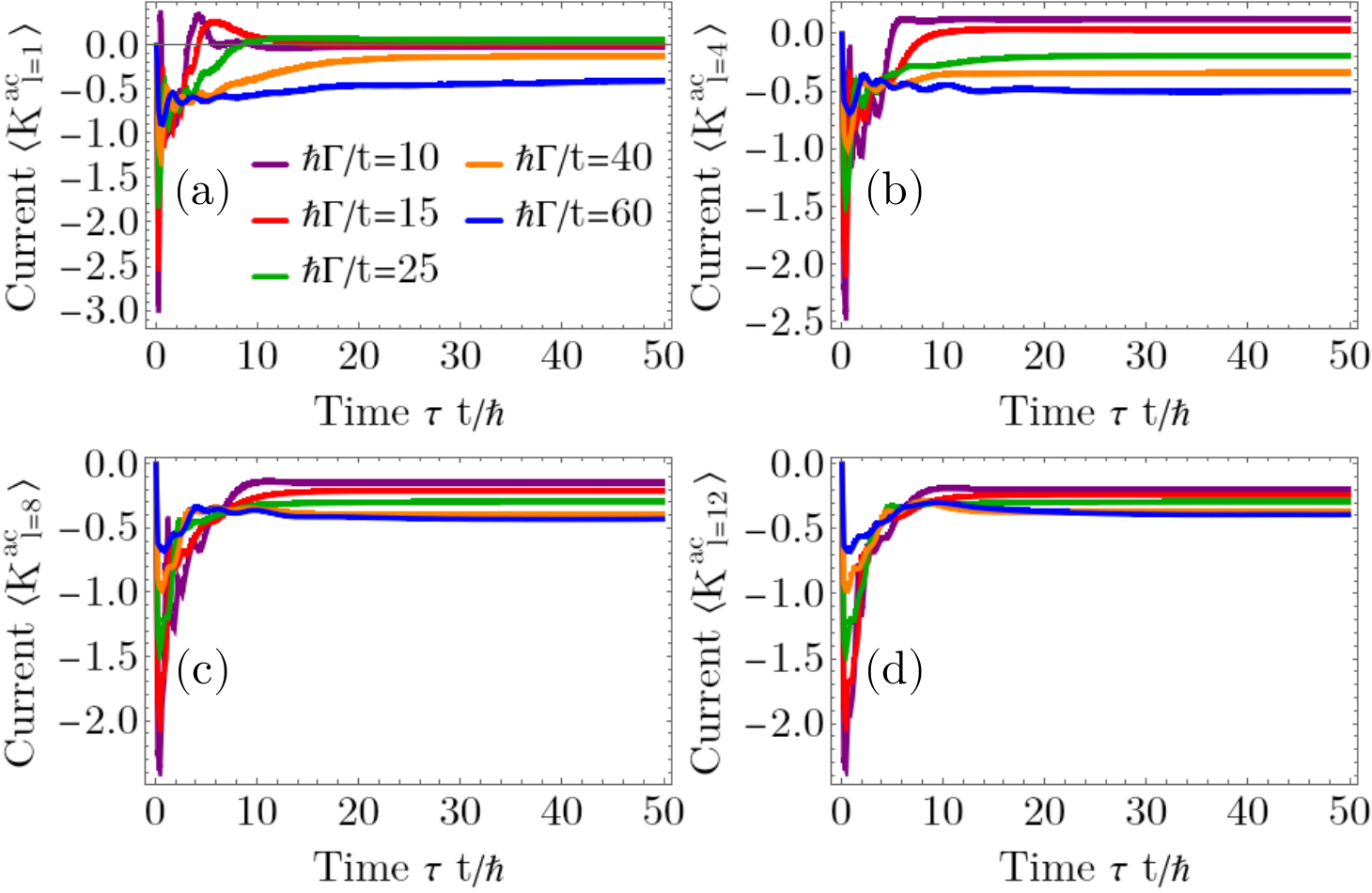}
\caption{The time dependence of the cavity contribution to the current, $\langle\hat{K}^\text{ac}_l\rangle$, for the global atom-cavity model, Eqs.~(\ref{eq:Lindblad})-(\ref{eq:Hamiltonian_ac}), and different values of the dissipation strength, $\hbar\Gamma/t\in\{10,15,25,40,60\}$. 
The different panels correspond to the different sites for which the $\langle\hat{K}^\text{ac}_l\rangle$ was computed, $l\in\{1,4,8,12\}$.
The other parameters are $L=32$ sites, $N=8$ particles, detuning $\hbar\delta/t=5$ and atoms-cavity coupling strength $\hbar\Omega/t=2.5$.
}
\label{fig:current_ac_1}
\end{figure}

\begin{figure}[!hbtp]
\centering
\includegraphics[width=0.48\textwidth]{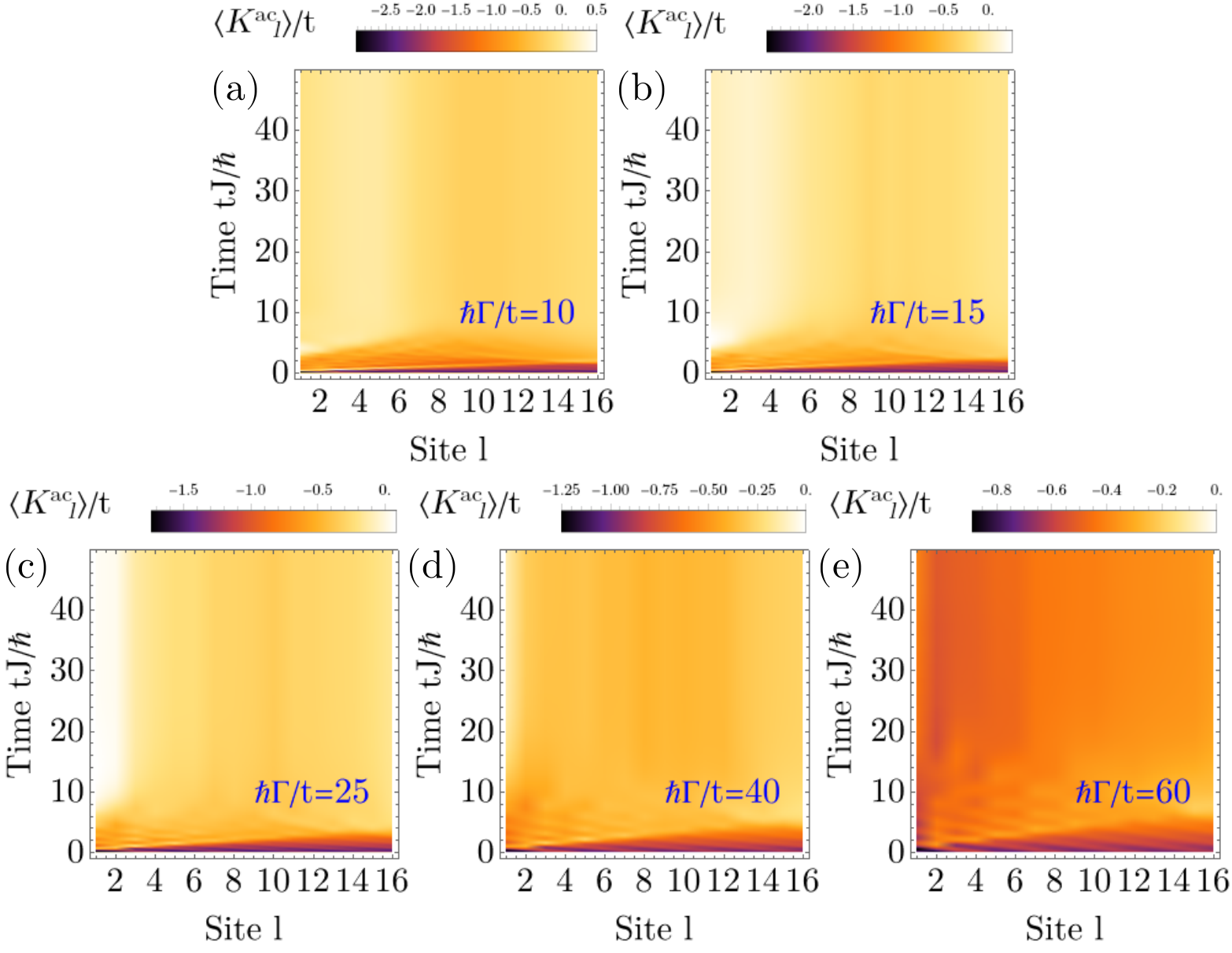}
\caption{The time dependence of the cavity contribution to the current, $\langle\hat{K}^\text{ac}_l\rangle$, for the first half of the chain, $1\leq l \leq 16$, for the global atom-cavity model, Eqs.~(\ref{eq:Lindblad})-(\ref{eq:Hamiltonian_ac}). 
The different panels correspond to different values of the dissipation strength, $\hbar\Gamma/t\in\{10,15,25,40,60\}$.
The other parameters are $L=32$ sites, $N=8$ particles, detuning $\hbar\delta/t=5$ and atoms-cavity coupling strength $\hbar\Omega/t=2.5$.
}
\label{fig:current_ac_2}
\end{figure}

\begin{figure}[!hbtp]
\centering
\includegraphics[width=0.4\textwidth]{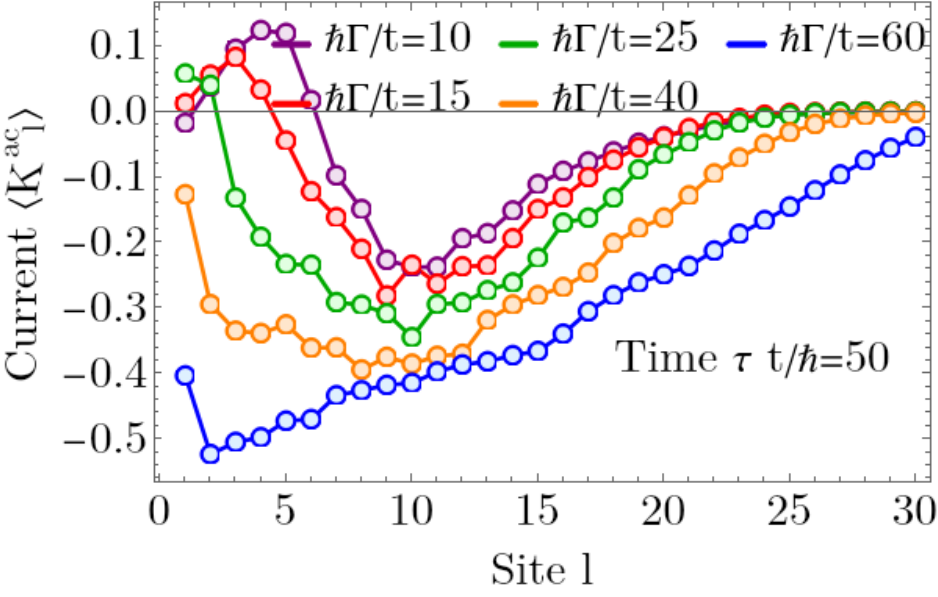}
\caption{The space dependence of the cavity contribution to the current, $\langle\hat{K}^\text{ac}_l\rangle$ at time $\tau t/\hbar=50$, for the global atom-cavity model, Eqs.~(\ref{eq:Lindblad})-(\ref{eq:Hamiltonian_ac}), and different values of the dissipation strength, $\hbar\Gamma/t\in\{10,15,25,40,60\}$. 
The other parameters are $L=32$ sites, $N=8$ particles, detuning $\hbar\delta/t=5$ and atoms-cavity coupling strength $\hbar\Omega/t=2.5$.
}
\label{fig:current_ac_3}
\end{figure}

We show the dynamics of the contributions to the currents stemming from the atomic kinetic energy, $\langle\hat{J}_l\rangle$, for the atom-cavity model in Fig.~\ref{fig:current_cav_1} and Fig.~\ref{fig:current_cav_2}. We observe that the onset of the currents depends both on the position in the chain and the value of the dissipation strength $\Gamma$. $\langle\hat{J}_l\rangle$ increases first for smaller $l$, with the propagation front depending approximately linearly on the site position $l$ and with the slope of this front decreasing with the value of $\hbar\Gamma/t$ (see Fig.~\ref{fig:current_cav_1} and Fig.~\ref{fig:current_cav_2}).
In contrast to the behavior of the currents $\langle\hat{J}_l\rangle$ in the local atomic model with kinetic dissipation (Fig.~\ref{fig:current_kin_1} and Fig.~\ref{fig:current_kin_2}), where after the initial increase the currents remained finite only for a few sites, in the atom-cavity model we obtain that the currents $\langle\hat{J}_l\rangle$ have a finite magnitude for most of the sites of the atomic chain (see Fig.~\ref{fig:current_cav_1} and Fig.~\ref{fig:current_cav_2}).
The site dependence of $\langle\hat{J}_l\rangle$ at long times is shown in Fig.~\ref{fig:current_cav_3}. Interestingly, the spatial region with large values of the currents and the magnitude of these values increases as we increase the dissipation strength $\Gamma$.
This signals that the decoherence free subspace can indeed host states with local atomic currents throughout the chain.

\begin{figure}[!hbtp]
\centering
\includegraphics[width=0.4\textwidth]{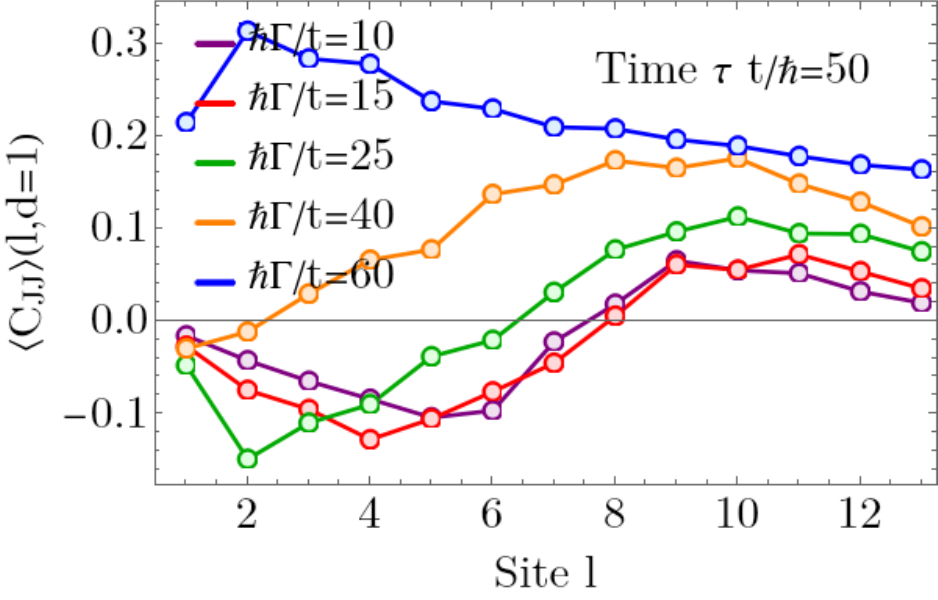}
\caption{The space dependence of the current-current correlations, $\langle\hat{C}_{JJ}\rangle(l,d)$ at time $\tau t/\hbar=50$ and distance $d=1$, for the global atom-cavity model, Eqs.~(\ref{eq:Lindblad})-(\ref{eq:Hamiltonian_ac}), and different values of the dissipation strength, $\hbar\Gamma/t\in\{10,15,25,40,60\}$. 
The other parameters are $L=32$ sites, $N=8$ particles, detuning $\hbar\delta/t=5$ and atoms-cavity coupling strength $\hbar\Omega/t=2.5$.
}
\label{fig:current_corr_cav_1}
\end{figure}

\begin{figure}[!hbtp]
\centering
\includegraphics[width=0.48\textwidth]{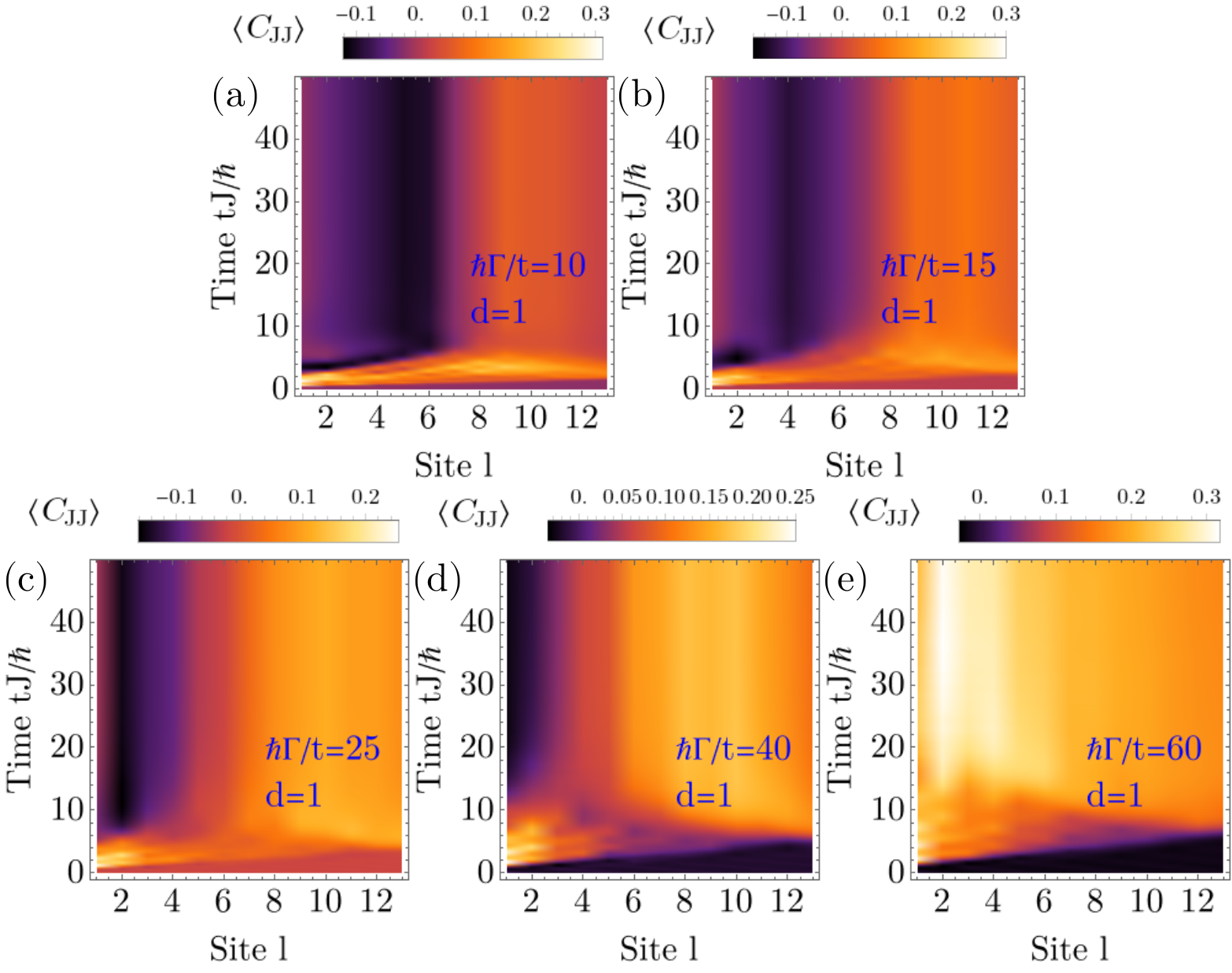}
\caption{The time dependence of the current-current correlations, $\langle\hat{C}_{JJ}\rangle(l,d)$, for distance $d=1$ and the sites $1\leq l \leq 14$, for the global atom-cavity model, Eqs.~(\ref{eq:Lindblad})-(\ref{eq:Hamiltonian_ac}).
The different panels correspond to the different values of the dissipation strength, $\hbar\Gamma/t\in\{10,15,25,40,60\}$.
The other parameters are $L=32$ sites, $N=8$ particles, detuning $\hbar\delta/t=5$ and atoms-cavity coupling strength $\hbar\Omega/t=2.5$.
}
\label{fig:current_corr_cav_2}
\end{figure}

As the time-derivative of the local densities, Eq.~(\ref{eq:continuity_kin_cav}), is vanishing in the steady state of the system, this implies that the finite values of the currents $\langle\hat{J}_l\rangle$ need to be compensated by finite values of the contributions to the currents from the coupling to the cavity, $\langle\hat{K}^\text{ac}_l\rangle$, Eq.~(\ref{eq:current_kin_cav}).
We plot the dynamics of $\langle\hat{K}^\text{ac}_l\rangle$ in Fig.~\ref{fig:current_ac_1} and Fig.~\ref{fig:current_ac_2}. 
We observe a rapid increase at short time followed by a decay towards a steady value for all the values of $\Gamma$ and distances considered.
The initial peak can be attributed to the large increase in the photon number at short times, as seen in Fig.~\ref{fig:kinetic_energy_photon_no}(b), which couples to the finite expectation value of the kinetic energy in the initial state.
The value at long times to which $\langle\hat{K}^\text{ac}_l\rangle$ evolves compensates the one of $\langle\hat{J}_l\rangle$, as seen when comparing Fig.~\ref{fig:current_ac_3} to Fig.~\ref{fig:current_cav_3}.
This implies that atomic kinetic terms in the Hamiltonian generate a particle current in one direction, while the cavity induces particle currents in the opposite direction.

\begin{figure}[!hbtp]
\centering
\includegraphics[width=0.48\textwidth]{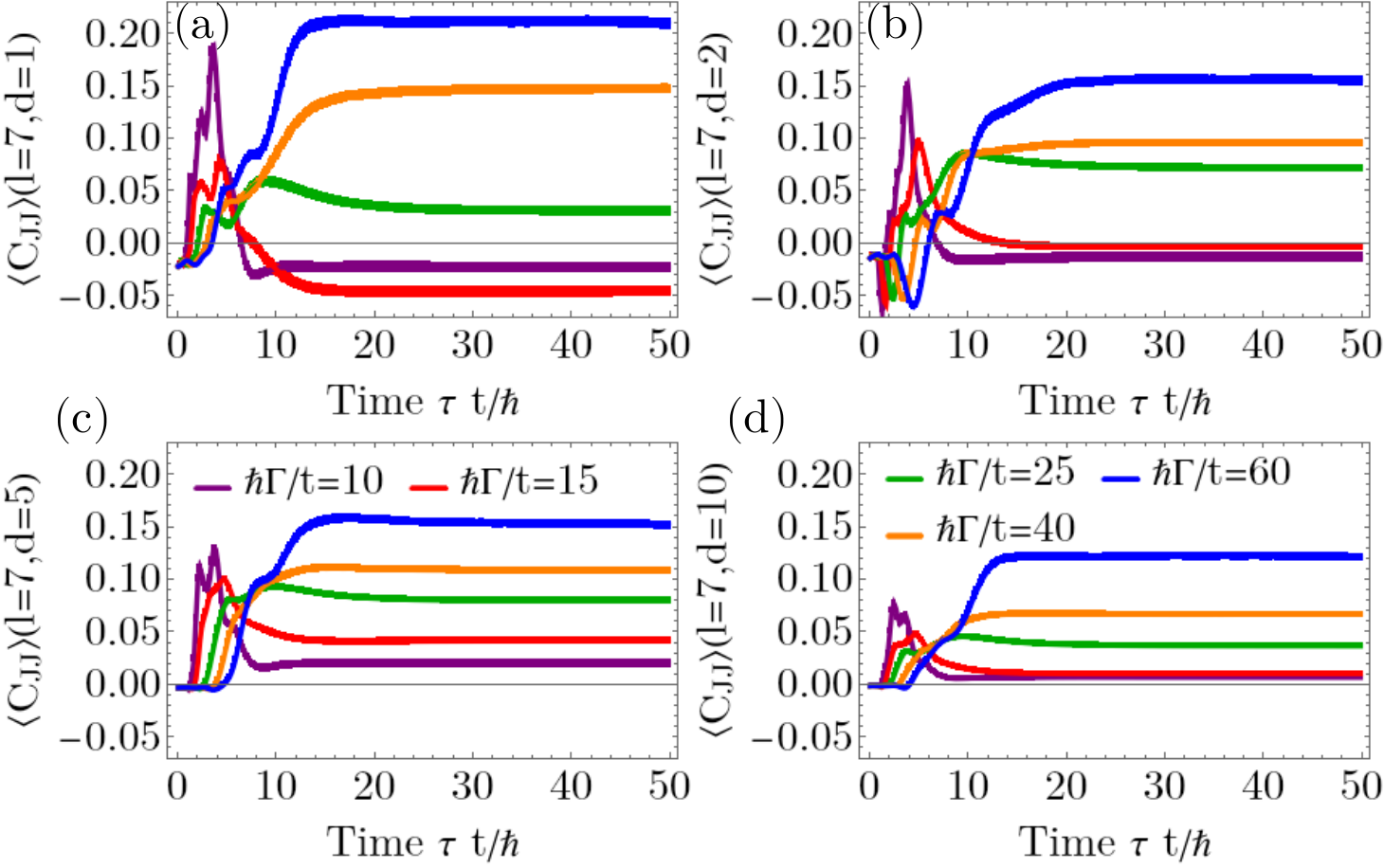}
\caption{The time dependence of the current-current correlations, $\langle\hat{C}_{JJ}\rangle(l,d)$ for the site $l=7$, for the global atom-cavity model, Eqs.~(\ref{eq:Lindblad})-(\ref{eq:Hamiltonian_ac}), and different values of the dissipation strength, $\hbar\Gamma/t\in\{10,15,25,40,60\}$. 
The different panels correspond to the different distances for which the $\hat{C}_{JJ}(l,d)$ was computed, $d\in\{1,2,5,10\}$.
The other parameters are $L=32$ sites, $N=8$ particles, detuning $\hbar\delta/t=5$ and atoms-cavity coupling strength $\hbar\Omega/t=2.5$.
}
\label{fig:current_corr_cav_3}
\end{figure}

\begin{figure}[!hbtp]
\centering
\includegraphics[width=0.4\textwidth]{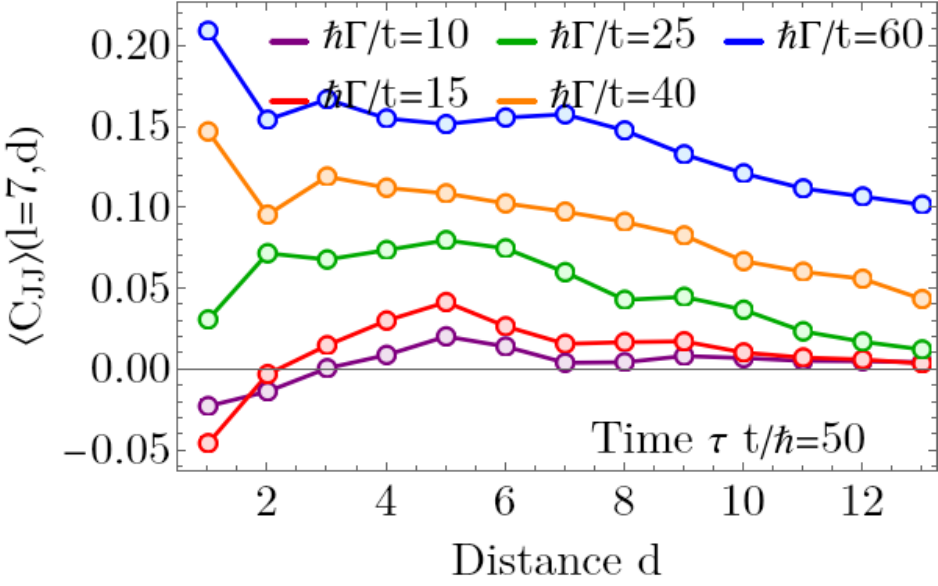}
\caption{The distance dependence of the current-current correlations, $\langle\hat{C}_{JJ}\rangle(l,d)$ at time $\tau t/\hbar=50$ and the site $l=7$, for the global atom-cavity model, Eqs.~(\ref{eq:Lindblad})-(\ref{eq:Hamiltonian_ac}), and different values of the dissipation strength, $\hbar\Gamma/t\in\{10,15,25,40,60\}$. 
The other parameters are $L=32$ sites, $N=8$ particles, detuning $\hbar\delta/t=5$ and atoms-cavity coupling strength $\hbar\Omega/t=2.5$.
}
\label{fig:current_corr_cav_4}
\end{figure}

\begin{figure}[!hbtp]
\centering
\includegraphics[width=0.48\textwidth]{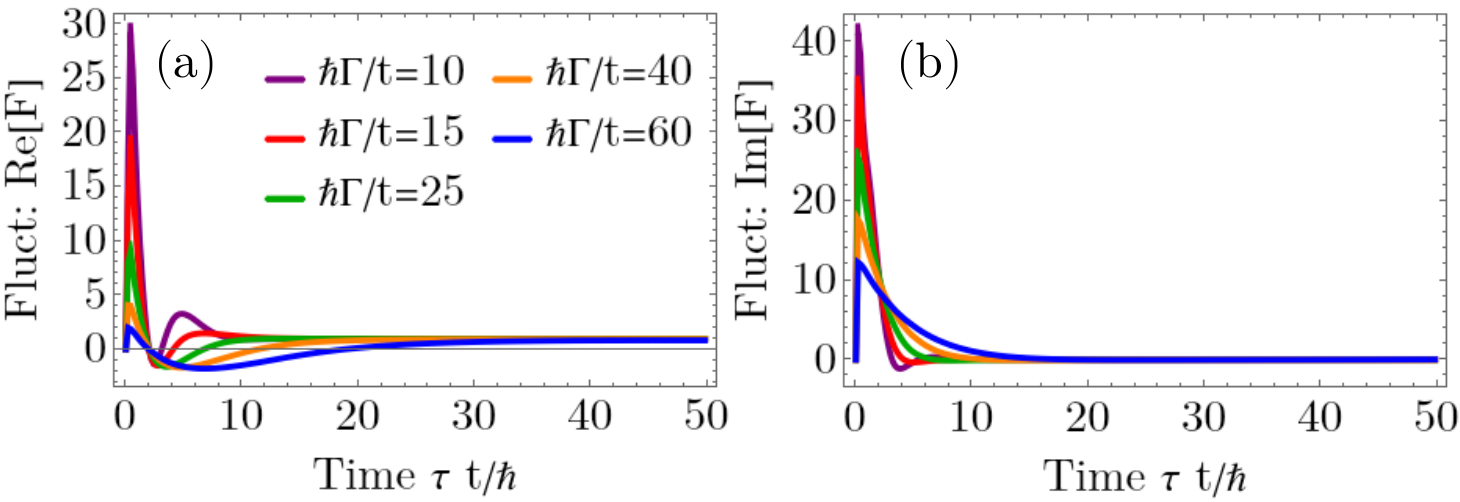}
\caption{The time dependence of the fluctuations of the (a) real part, (b) imaginary part of the fluctuations of the atoms-cavity coupling $F$, Eq.~(\ref{eq:fluct}), for different values of the dissipation strength, $\hbar\Gamma/t\in\{10,15,25,40,60\}$. 
The other parameters are $L=32$ sites, $N=8$ particles, detuning $\hbar\delta/t=5$ and atoms-cavity coupling strength $\hbar\Omega/t=2.5$.
}
\label{fig:fluct_cav}
\end{figure}

We can now tackle one of the main questions posed by this study: can we also stabilize current-current correlations at long times, and in particular long-range correlations?
We saw in Sec.~\ref{sec:results_atoms} that for the atomic chain coupled to kinetic dissipation we can stabilize finite current-current correlations, $\langle\hat{C}_{JJ}\rangle(l,d)$, only in a narrow spatial region and that they are rapidly decaying with the distance $d$.
We contrast this with the results for $\langle\hat{C}_{JJ}\rangle(l,d)$ in the global atom-cavity model, Eqs.~(\ref{eq:Lindblad})-(\ref{eq:Hamiltonian_ac}).
We first analyze the influence of the starting site for the correlations at distance $d=1$, for which we depict the value at $\tau t/\hbar=50$ in Fig.~\ref{fig:current_corr_cav_1} and the time dynamics in Fig.~\ref{fig:current_corr_cav_2}.
As for the local currents, we obtain finite values for $\langle\hat{C}_{JJ}\rangle(l,d=1)$ throughout the chain. 
For $\hbar\Gamma/t=10$ and $\hbar\Gamma/t=15$ we observe smaller values and a chain of sign around $l=8$, however, the correlations become stronger as we increase the dissipation strength, with the largest value for the considered parameters and the spatial region shown obtained for $\hbar\Gamma/t=60$.
Thus, the coupling to the cavity can indeed induce finite current-current correlations, $\langle\hat{C}_{JJ}\rangle(l,d)$.
In order to understand the behavior at longer distances, in Fig.~\ref{fig:current_corr_cav_3} we show the dynamics of $\langle\hat{C}_{JJ}\rangle(l=7,d)$ for different distances $d$.
As for $\hbar\Gamma/t=10$ and $\hbar\Gamma/t=15$ $\langle\hat{C}_{JJ}\rangle(l=7,d)$ has a relatively small value already at $d=1$, we focus on the results obtained for $\hbar\Gamma/t\geq25$.
We observe that after the initial increase for all the distances shown, including $d=10\approx L/3$, $\langle\hat{C}_{JJ}\rangle(l=7,d)$ has finite values at longer times, which, importantly, are increasing with the value of $\Gamma$.
The late time values of $\langle\hat{C}_{JJ}\rangle(l=7,d)$ are shown in Fig.~\ref{fig:current_corr_cav_4}.
Here we observe both the finite values of $\langle\hat{C}_{JJ}\rangle(l=7,d)$ for large values of the dissipation strength and the fact that the correlations are only slowly decaying with the distance $d$.
This is in sharp contrast with the rapid decay shown in Fig.~\ref{fig:current_corr_kin_4} for atoms under kinetic dissipation.

One interesting question that can be asked in hybrid atom-cavity systems relates to the role of the quantum fluctuations in the coupling of the atoms to the cavity \cite{BezvershenkoRosch2021, HalatiKollath2025, HalatiJager2025, SchmitJager2026}. To compute this we compute the following quantity
\begin{align}
\label{eq:fluct}
F\equiv\sum_l\left(\langle\hat{a}\hat{b}_{l} \hat{b}_{l+1}^\dagger\rangle-\langle\hat{a}\rangle\langle\hat{b}_{l} \hat{b}_{l+1}^\dagger\rangle\right),
\end{align} 
which quantifies the fluctuations in the coupling. This quantity would be zero if by performing the mean-field decoupling of the atoms and the cavity field one could recover the exact dynamics \cite{RitschEsslinger2013, MivehvarRitsch2021}.
We plot the time dependence of the real and imaginary part of $F$ in Fig.~\ref{fig:fluct_cav}. We observe a very large increase at short times, related to the non-equilibrium dynamics of coupling the atoms and the cavity by a quench. 
This is followed by an intermediate regime in which both the real and imaginary part of $F$ are finite. However, at long times $\Re[F]$ saturates to a finite value, marking the importance of fluctuations in the steady state of the coupled system, while $\Im[F]$ vanishes.
The fact that $\Im[F]=0$ can help us understand the dynamics of the cavity induced currents $\langle\hat{K}^\text{ac}_l\rangle \propto \Im\langle\hat{a}^\dagger\hat{b}_{l}^\dagger \hat{b}_{l+1}\rangle$, Eq.~(\ref{eq:current_kin_cav_av}).
As $\Im[F]=0$ and $\Im\langle a \rangle=0$ (not shown) in the steady state for the considered parameters we obtain $\langle\hat{K}^\text{ac}_l\rangle \propto \Re\langle\hat{a}^\dagger\rangle\Im\langle\hat{b}_{l}^\dagger \hat{b}_{l+1}\rangle\propto\Re\langle\hat{a}^\dagger\rangle\langle\hat{J}_l\rangle$, relating the currents stemming from the kinetic terms and the coupling to the cavity. 
Furthermore, this implies a value of the cavity field of $\Re\langle\hat{a}\rangle=-\frac{t}{2\hbar\Omega}$ in the steady state, interestingly independent of the dissipation strength, which is consistent with the value obtained in our results for all values of $\Gamma$.
Interestingly, the cavity field depends inverse proportionally with the coupling strength. This dependence is possible due to existence of multiple steady state for the atom-cavity system.
The study of the influence of the initial states on the dynamics and long-time properties and the analysis of the decoherence free subspace spanned by the operator $\mathcal{O}=\sum_{j=1}^{L-1}  \hat{b}_{j}^\dagger \hat{b}_{j+1}$ is left for future studies.

To summarize the results presented in this section, we obtained that for a chain of hardcore bosons, which are globally coupled to the field of a dissipative cavity by tunneling terms, strong local currents are obtained throughout the chain when the cavity photon losses have a large value. Furthermore, the atoms also exhibit finite current-current correlations which are only slowly decaying with the distance. This shows the complex coherent quantum dynamics that can emerge from the global coupling to a dissipative environment.

\section{Discussions and Conclusions \label{sec:conclusions}}

To conclude, we investigated the out-of-equilibrium dynamics of strongly interacting bosonic atoms confined to a one-dimensional chain in two different scenarios exhibiting nonreciprocal dissipative couplings.
In the local atomic model, Eqs.~(\ref{eq:Lindblad_atoms})-(\ref{eq:Hamiltonian_BH}), the dissipative processes act \emph{locally} on the atomic chain in the form of tunneling terms. This realizes a setup in each bond connecting two neighboring sites is coupled to a distinct \emph{Markovian} environment which moves particles in a certain direction.
In the atom-cavity model, Eqs.~(\ref{eq:Lindblad})-(\ref{eq:Hamiltonian_ac}), the atoms are \emph{globally} coupled to the field of an optical cavity via tunneling terms. As the photons in the cavity experience losses, the atoms effectively experience the action of a nonreciprocal \emph{non-local} environment inducing a preferred directionality for their motion. Furthermore, for cavity energy scales away from the adiabatic elimination limit the photon field realizes an effective \emph{non-Markovian} reservoir.
We have computed the numerically exact dissipative dynamics of the quantum correlations for both models by employing matrix product states methods.
We focused on the question on how to induce a directional motion for the atoms, manifested by finite values of the current operators, while maintaining their quantum coherence, as captured by the current-current correlations.
For the local atomic model with kinetic dissipation, Eqs.~(\ref{eq:Lindblad_atoms})-(\ref{eq:Hamiltonian_BH}), for strong dissipation the atoms accumulate at one end of the chain forming a rather sharp interface between the filled and empty sites. We obtain finite values of the currents only around this interface, with the contributions stemming from the Hamiltonian kinetic terms and the dissipative processes canceling each other. This can visualized in a semiclassical picture in which the Hamiltonian tunneling moves the atoms away from the interface, but the kinetic dissipation brings them back.
Finite values of the current-current correlations are also restricted to the narrow region around the interface and decay rapidly with the distance between the sites.
In contrast, for the atom-cavity model, Eqs.~(\ref{eq:Lindblad})-(\ref{eq:Hamiltonian_ac}), we obtain finite values of the currents throughout the chain, with their values increasing as the photon losses are stronger.
In this scenario, the contributions to the currents, which cancel in the steady state, either stem from the tunneling terms, or the cavity-assisted tunneling, where the atoms move by creating, or annihilating, photons.
When the atoms are coupled to the cavity we do obtain finite values for the current-current correlations in extended regions of the systems and they only slowly decay with the distance.
We attribute the contrasting behavior of the two models considered to the different nature of their decoherence free subspaces, as they either spanned by a large number of local operators, or by a single global operator. 
This allows for the coupling to the cavity to induce the current-current correlations and to protect them at long times.
Thus, by coupling the interacting atoms to the field of the optical cavity we obtain a platform capable of exhibiting complex long-range quantum correlations.
One can further envision the exciting prospect for studying multi-time correlations in such hybrid dissipative systems \cite{BuchholdDiehl2015, SciollaKollath2015, HalimehPiazza2018, WolffKollath2019}, in order to investigate the interplay of non-locality and non-reciprocity. 

The setups investigated in this work are directly relevant for experiments in which ultracold atoms trapped in optical lattices are coupled to the field of  optical cavities \cite{KlinderHemmerich2015b, LandigEsslinger2016, HrubyEsslinger2018}. 
New generations of experiments, which combine the optical cavities with microscopes \cite{OrsiBrantut2024, BologniniBrantut2025}, or tweezer arrays \cite{HartungRempe2024, HoStamper-Kurn2025, SeubertDistante2025}, are bringing the detection of complex long-range correlations within reach.
The dynamics of currents has been studied in other experimental setups employing ultracold atoms, either in transport measurements between reservoirs \cite{HusmannBrantut2015, HuangEsslinger2023}, or in the context of artificial magnetic fields \cite{AtalaBloch2014} and the Hall effect \cite{ZhouFallani2023, ZhouFallani2025}. 
Recently developed experimental schemes allow access even to the local currents stemming from tunneling terms \cite{ImpertroAidelsburger2024, ImpertroAidelsburger2025}.
Furthermore, while in this work we concentrated in setups relevant for ultracold atoms in optical cavities, we expect that our results can have a wider relevance for other hybrid quantum systems in which interacting particles are coupled to quantum bosonic fields, e.g.~trapped ions coupled to phonons \cite{TomzaJulienne2019, MonroeYao2021}, superconducting circuits \cite{LEHURSchiro2016, BlaisWallraff2021}, or electrons in solid state cavities \cite{CongKono2016, FriskKockumNori2019, FornDiazSolano2019, SchlawinSentef2022}.

\section*{DATA AVAILABILITY}

The supporting data for this article are openly available at Zenodo \cite{datazenodo}.

\section*{ACKNOWLEDGMENTS}

I thank J.-S.~Bernier, T.~Giamarchi, L.~Pizzino, and S.~Sur for fruitful discussions.
I acknowledge support in part by the Swiss National Science Foundation under Division II Grant No.~200020-219400.

\end{document}